\documentclass[conf]{new-aiaa}
\usepackage[utf8]{inputenc}
\usepackage{booktabs}
\usepackage{multirow}
\usepackage{graphicx}
\usepackage{wrapfig}
\usepackage{subcaption}
\usepackage{amsmath}
\usepackage[version=4]{mhchem}
\usepackage{siunitx}
\usepackage{cleveref}
\usepackage{longtable,tabularx}
\makeatletter
\renewcommand\@makefntext[1]{\leftskip=2em \hskip0em \@makefnmark#1}
\makeatother

\title{Planetary Exploration 3.0: A Roadmap for Software-Defined, Radically Adaptive Space Systems}

\author{
Masahiro Ono$^{1}$,
Daniel Selva$^{2}$,
Morgan L. Cable$^{1,3,4}$,
Marie Ethvignot$^{1,5}$,
Margaret Hansen$^{6}$,
Andreas M. Hein$^{7}$,
Elena-Sorina Lupu$^{1,8,9}$,
Zachary Manchester$^{10}$,
David Murrow$^{11}$,
Chad Pozarycki$^{12}$,
Pascal Spino$^{10}$,
Amanda Stockton$^{13}$,
Mathieu Choukroun$^{1}$,
Soon-Jo Chung$^{1,14}$,
John Day$^{15}$,
Alexander Demagall$^{2}$,
Anthony Freeman$^{1}$,
Chloe Gentgen$^{10}$,
Michel D. Ingham$^{1}$,
Charity M. Phillips-Lander$^{16}$,
Richard Rieber$^{1}$,
Alejandro Salado$^{17}$,
Maria Sakovsky$^{18}$,
Lori R. Shiraishi$^{1}$,
Yisong Yue$^{14}$,
Kris Zacny$^{19}$
}

\affil{
$^{1}$Jet Propulsion Laboratory, California Institute of Technology, Pasadena, CA 91109, USA \\
$^{2}$Texas A\&M University, College Station, TX 77840, USA \\
$^{3}$Planetary Science Institute, AZ 85719, USA \\
$^{4}$Te Herenga Waka -- Victoria University of Wellington, 6011, New Zealand \\
$^{5}$Swiss Federal Institute of Technology, Lausanne 1015, Switzerland \\
$^{6}$Carnegie Mellon University, Pittsburgh, PA 15213, USA \\
$^{7}$SnT, University of Luxembourg, 1855, Luxembourg \\
$^{8}$Openmind Research Institute, Edmonton, Canada \\
$^{9}$California Institute of Technology, Pasadena, CA 91109, USA \\
$^{10}$Massachusetts Institute of Technology, Cambridge, MA 02139, USA \\
$^{11}$Space Connections, LLC, Highlands Ranch, CO 80126, USA \\
$^{12}$Deleon Technologies, Inc., Atlanta, GA 30309, USA \\
$^{13}$Georgia Institute of Technology, Atlanta, GA 30330, USA \\
$^{14}$California Institute of Technology, Pasadena, CA 91125, USA \\
$^{15}$Blue Origin, Kent, WA, USA \\
$^{16}$Space Science Division, Southwest Research Institute, San Antonio, TX 78238, USA \\
$^{17}$The University of Arizona, Tucson, AZ 85721, USA \\
$^{18}$Stanford University, Stanford, CA 94305, USA \\
$^{19}$Honeybee Robotics, a Blue Origin Company, Altadena, CA 91001, USA \\
}

\begin{document}

\maketitle

\begin{abstract}
The surface and subsurface of worlds beyond Mars remain largely unexplored. Yet these worlds hold keys to fundamental questions in planetary science---from potentially habitable subsurface oceans on icy moons to ancient records preserved in Kuiper Belt objects. NASA's success in Mars exploration was achieved through incrementalism: 22 progressively sophisticated missions over decades. This paradigm, which we call \textit{Planetary Exploration 2.0 (PE~2.0)}, is untenable for the outer Solar System, where cruise times of a decade or more make iterative missions infeasible. We propose \textit{Planetary Exploration 3.0 (PE~3.0)}: a paradigm in which unvisited worlds are explored by a single or a few missions with radically adaptive space systems. A PE~3.0 mission conducts both initial exploratory science and follow-on hypothesis-driven science based on its own \textit{in situ} data returns, evolving spacecraft capabilities to work resiliently in previously unseen environments. The key enabler of PE~3.0 is \textit{software-defined space systems} (SDSSs)---systems that can adapt their functions at all levels through software updates. This paper presents findings from a Keck Institute for Space Studies (KISS) workshop on PE~3.0, covering: (1) PE~3.0 systems engineering including science definition, architecture, design methods, and verification \& validation; (2) software-defined space system technologies including reconfigurable hardware, multi-functionality, and modularity; (3) onboard intelligence including autonomous science, navigation, controls, and embodied AI; and (4) three PE~3.0 mission concepts: a Neptune/Triton smart flyby, an ocean world explorer, and an Oort cloud reconnaissance mission.

\end{abstract}

\section{Introduction: Planetary Exploration 3.0}\label{sec:intro}
\subsection{The Vision and the Goals}
\begin{figure}[tb]
    \centering
    
    \begin{subfigure}[b]{0.33\textwidth}
        \centering
        \includegraphics[width=\textwidth]{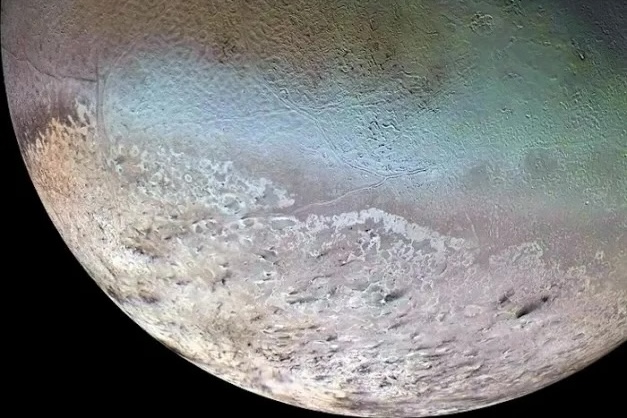}
        \caption{}
    \end{subfigure}
    \hfill
    \begin{subfigure}[b]{0.33\textwidth}
        \centering
        \includegraphics[width=\textwidth]{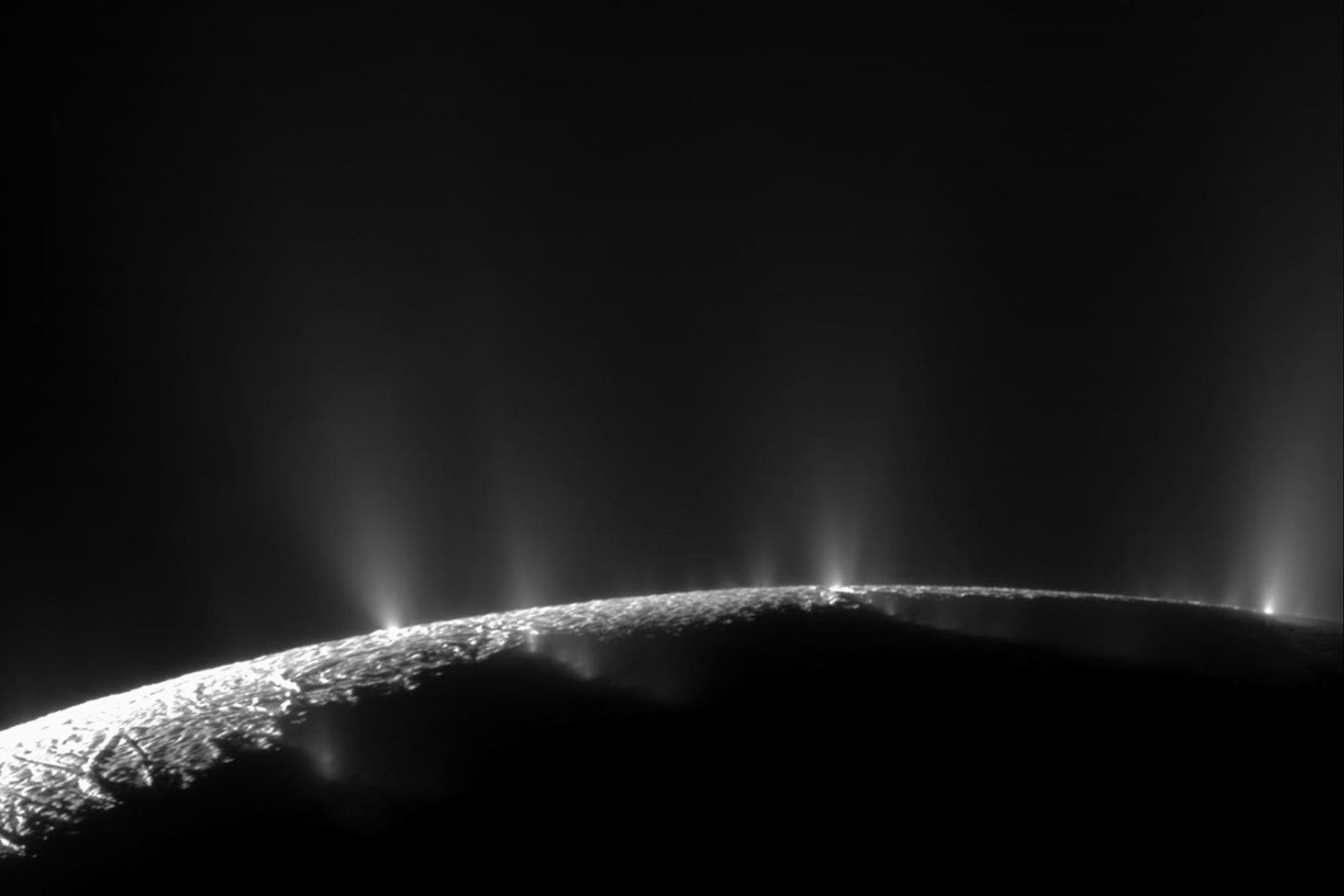}
        \caption{}
    \end{subfigure}
    \hfill
    \begin{subfigure}[b]{0.33\textwidth}
        \centering
        \includegraphics[width=\textwidth]{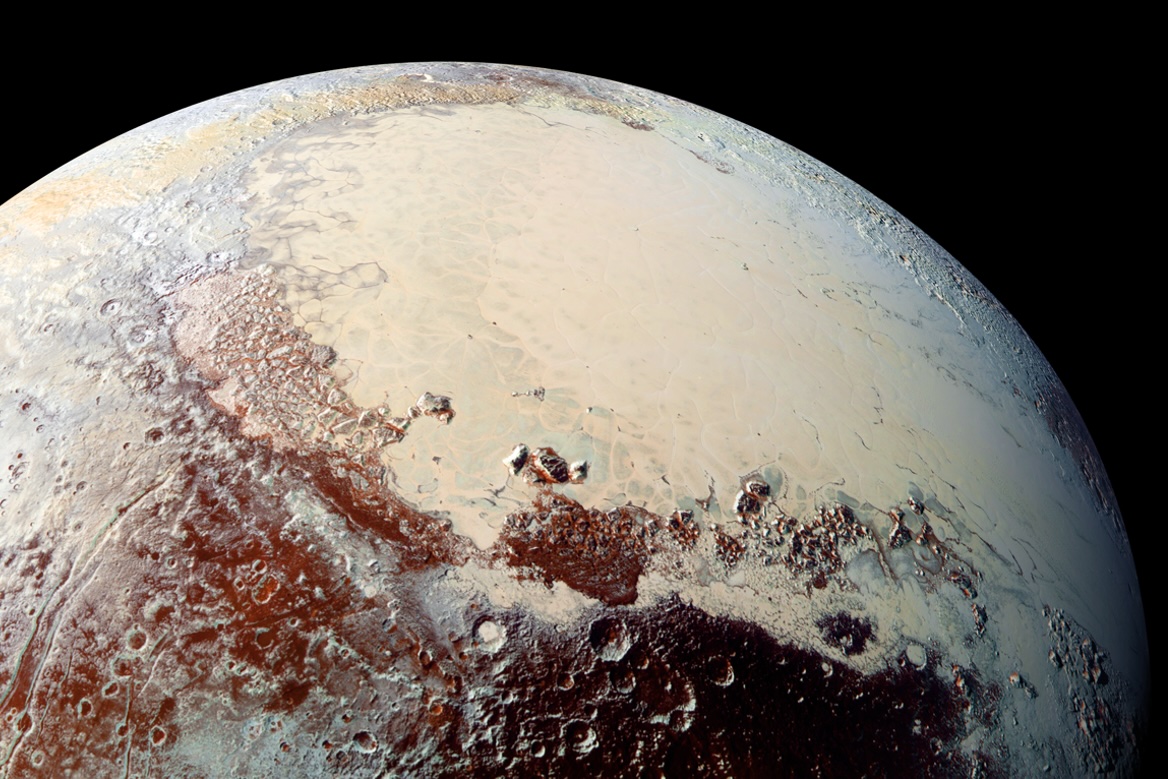}
        \caption{}
    \end{subfigure}
    
    \vspace{0.5em}
    
    \begin{subfigure}[b]{0.33\textwidth}
        \centering
        \includegraphics[width=\textwidth]{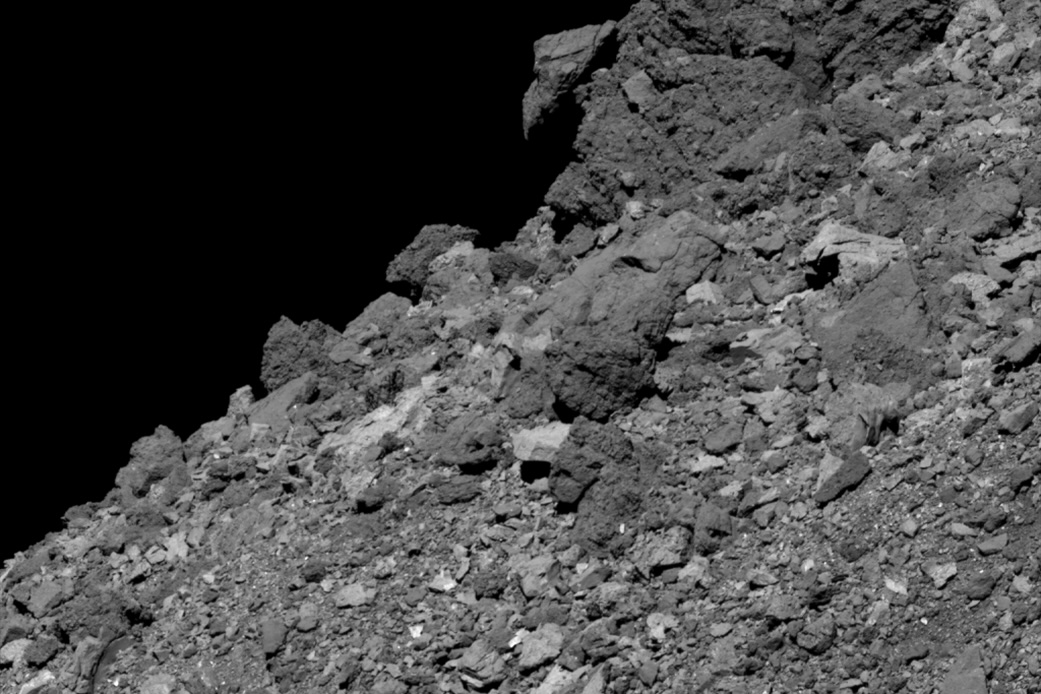}
        \caption{}
    \end{subfigure}
    \hfill
    \begin{subfigure}[b]{0.33\textwidth}
        \centering
        \includegraphics[width=\textwidth]{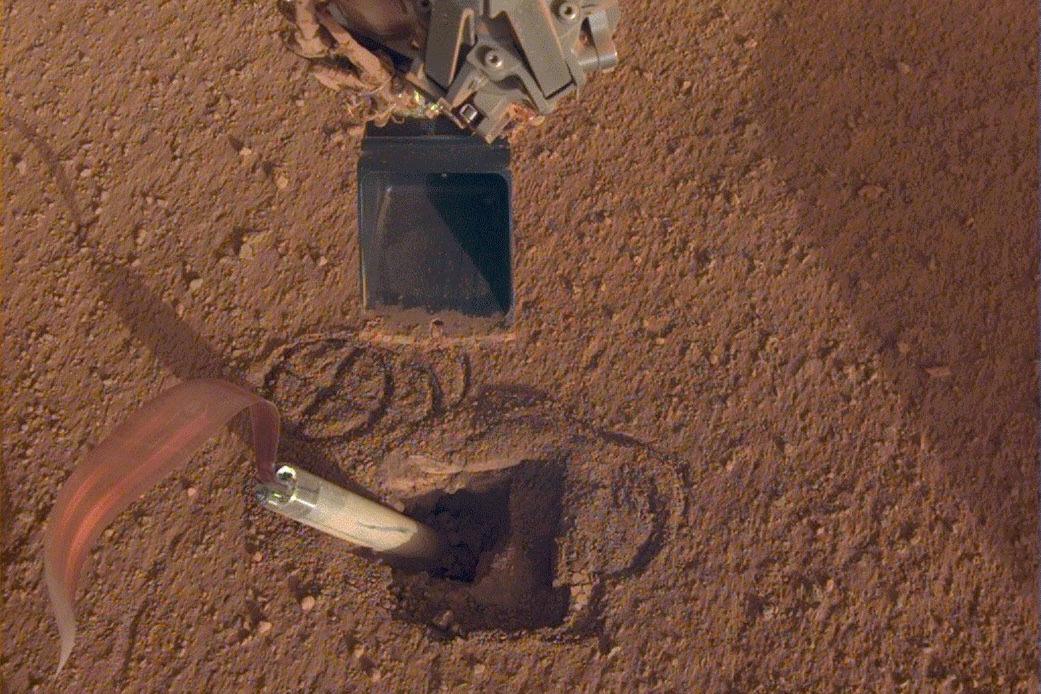}
        \caption{}
    \end{subfigure}
    \hfill
    \begin{subfigure}[b]{0.33\textwidth}
        \centering
        \includegraphics[width=\textwidth]{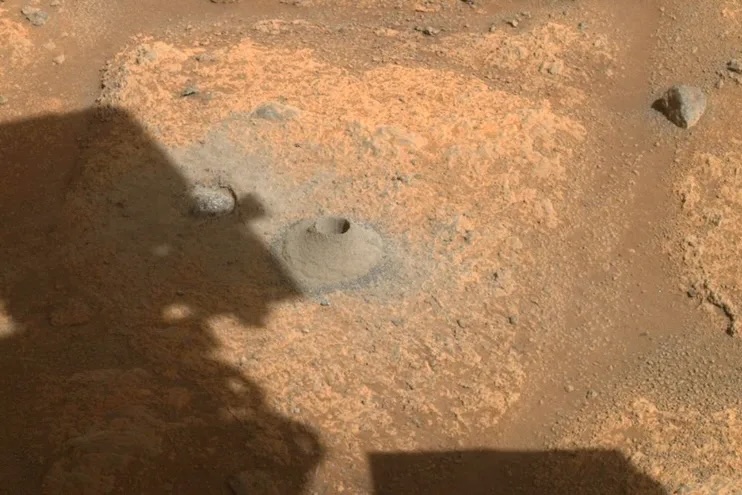}
        \caption{}
    \end{subfigure}
    
    \caption{Planetary exploration is characterized by unknowns, which are both the primary reason \textit{and} challenge for exploration. (a) The active nitrogen geysers on Triton, the largest moon of Neptune and likely a captured Kuiper belt object, discovered by Voyager 2. (b) The active geysers on Enceladus, an icy moon of Saturn, which discharge water from the potentially habitable subsurface ocean, discovered by Cassini. NASA/JPL-Caltech/SSI. (c) The heart-shaped Sputnik Planitia on Pluto, discovered by New Horizons, is indicative of ongoing geological activity, a major surprise to planetary scientists. NASA/JHUAPL/SWRI. (d) The unexpectedly rocky surface of Bennu posed a substantial challenge for OSIRIS-REx's touchdown navigation. NASA/Goddard/University of Arizona. (e) InSight’s heat probe mole failed to reach the target depth due to unexpected terrain characteristics. NASA/JPL-Caltech. (f) Perseverance’s first sample coring attempt failed since the rock was unexpectedly brittle, and it crumbled to powder. NASA/JPL-Caltech.}
    \label{fig:unknowns}
\end{figure}

 Space is full of unknowns, and that is why we explore. We have never visited new planets, moons, or small bodies without encountering surprises beyond our imaginations---active volcanoes on Io, plumes on Triton (\Cref{fig:unknowns}-a), and the evidence of ongoing geological activity on Pluto \cite{stern2015pluto} (\Cref{fig:unknowns}-c), just to name a few. 
Upon encountering such surprises, many missions often found themselves unprepared. 
For example, Cassini discovered the plume of Enceladus, which is believed to discharge material from the potentially habitable subsurface water ocean (Figure \ref{fig:unknowns}-b). In its extended mission, Cassini flew through the plume multiple times to analyze the chemical composition using its Ion and Neutral Mass Spectrometer (INMS) and Cosmic Dust Analyzer (CDA). However, these instruments were designed to characterize only low-mass organic and inorganic compounds because the astrobiological potential of Enceladus was not scientifically supported when Cassini’s science was defined. 
Unknown environments pose substantial engineering challenges as well. When OSIRIS-Rex first imaged Bennu, it found the asteroid's surface unexpectedly rocky (Figure \ref{fig:unknowns}-d). As a result, the LiDAR-based navigation planned to be used for touchdown was deemed too risky and was replaced by an experimental vision-based navigation approach \cite{lorenz2017lessons}. 
Even Mars, the most visited planet besides Earth, continues to surprise us. Two examples of many from recent missions include InSight’s heat probe mole, which failed to reach the target depth, likely due to unexpectedly low terrain cohesion (\Cref{fig:unknowns}-e), and Perseverance’s first sample coring attempt, which failed because the rock was surprisingly brittle, and it crumbled to powder (\Cref{fig:unknowns}-f). 

\begin{figure}[tb]
    \centering
    
    \begin{subfigure}[b]{\textwidth}
        \centering
        \includegraphics[width=\textwidth]{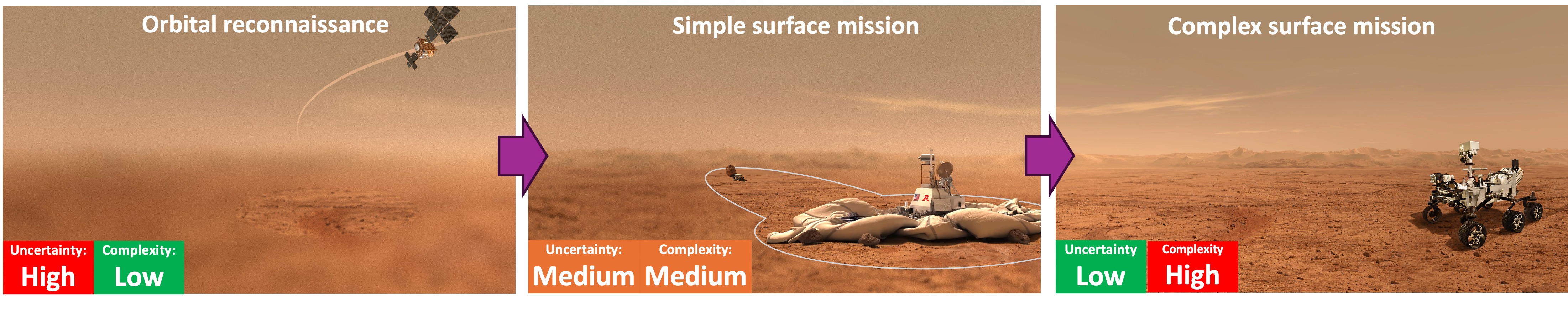}
        \caption{Planetary Exploration 2.0: Incremental sophistication over many missions, where the behaviors of the space system of each mission are largely pre-designed and fixed based on detailed environmental knowledge brought by prior missions.
}
    \end{subfigure}
    
    \vspace{0.5em}
    
    \begin{subfigure}[b]{\textwidth}
        \centering
        \includegraphics[width=\textwidth]{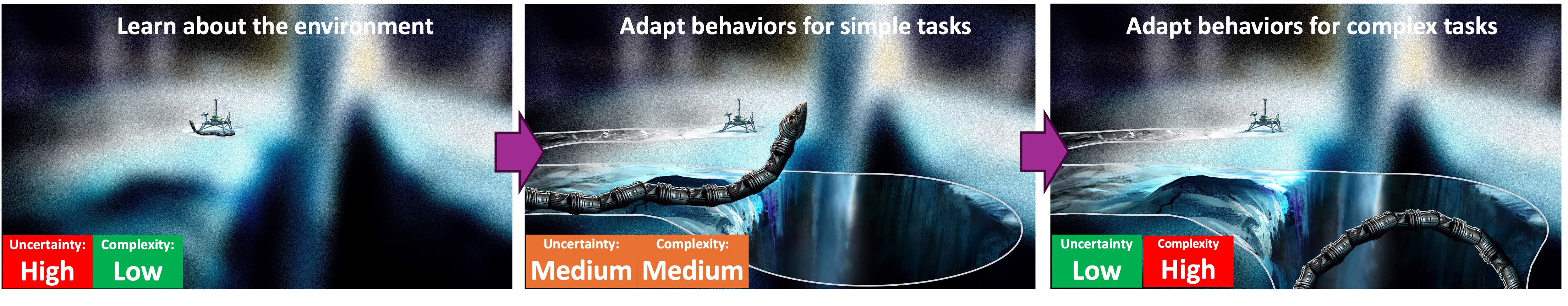}
        \caption{Planetary Exploration 3.0 (Proposed): One-shot exploration with a highly adaptive space system, which evolves its behaviors \textit{in situ} for incrementally complex tasks as it learns more about the environment.}
    \end{subfigure}
    
    \caption{Comparison of (a) the ongoing planetary exploration paradigm and (b) the proposed paradigm.}
    \label{fig:pe3}
\end{figure}

The conventional approach to addressing unknowns in planetary exploration has been through a campaign consisting of multiple incrementally sophisticated missions. 
As illustrated in~\Cref{fig:pe3}-a, such a campaign starts with low-risk exploratory missions, typically flybys or orbiters. Since much is unknown before these first missions, the scientific objectives are less specific and more exploratory while the task complexity tends to be low. The information gain of these early missions tends to be very high, resulting in major surprise discoveries and substantially improved constraints on environmental conditions. This allows scientists and engineers to define and design more specific and complex follow-on missions, which eventually reach the surface and provide intricate study of the \textit{in situ} environment. 
Finally, highly complex space systems are designed for in-depth exploration with highly specialized scientific payloads selected based on knowledge gleaned by the earlier missions.
As the environmental uncertainty is reduced, the scientific questions become deeper and more specific, making later missions increasingly more hypothesis-driven rather than exploratory. 
The space system design also benefits from the acquiredknowledge of the target environment, allowing the engineering requirements and the resulting design to be more specific and complex.
For the rest of this paper, we call this incremental exploration approach \textit{Planetary Exploration 2.0}, or \textit{PE 2.0} for short.

The most representative example of PE 2.0 is NASA's Mars exploration campaign, which so far consists of 22 missions spanning over seven decades. The program has advanced step-by-step from early exploratory missions, including flybys (e.g., Mariner 4 in 1964) and orbiters (Mariner 9 in 1971), to static landers (e.g., Viking in 1976), to simple rovers (e.g., Mars Pathfinder's Sojourner in 1997) and ultimately to the highly complex rovers Curiosity (2012) and Perseverance (2021). 
Countless discoveries from this highly successful campaign, including direct evidence of surface liquid water flow \cite{squyres2004situ,grotzinger2014habitable} and potential biosignatures from the ancient past \cite{hurowitz2025detection}, have completely reshaped our view of the red planet. 
In fact, Mars is perhaps the ``sweet spot" for the PE 2.0 model. 
The cruise time (6-8 months) and the frequency of launch opportunities (26 months) are too long for trial-and-error missions (or \textit{Planetary Exploration 1.0}), which worked well for pre-Apollo robotic exploration of the Moon, but short enough for incrementalism if a few decades of time is allowed. 

While the success of NASA's Mars exploration campaign is undeniable, the focus on the red planet in the past few decades left the surface and subsurface of worlds beyond Mars largely unexplored. Yet these worlds hold the keys to some of the greatest questions in planetary science and astrobiology. The icy moons of gas giants conceal potentially habitable subsurface oceans; Kuiper Belt objects preserve ancient records of the Solar System; and interstellar objects occasionally visit the Solar System, possibly carrying material from other star systems.
As we now set our sights on exploring a multitude of unvisited worlds in the outer Solar System, it is clear that the incremental PE 2.0 approach will be untenable. Long cruise times to outer Solar System destinations (e.g., $\sim$10 years to Saturn's moons) mean that following the Mars-style incremental approach would take centuries. Even if we could sustain support for centuries-long campaigns (which seems dubious), it is difficult to conceive reconnaissance missions that can adequately reduce risk for a subsequent subsurface exploration missions following this paradigm. 

Instead, we envision a new planetary exploration paradigm, \textit{Planetary Exploration 3.0 (PE~3.0)}~\cite{Ono2024}, in which unvisited worlds are explored in a single or a few missions.
The key enabler of PE 3.0 is radically adaptive space systems. 
As illustrated in \Cref{fig:pe3}-b, PE 3.0 replaces the incremental sophistication over multiple missions in PE 2.0 with the functional adaptability of a single space system.
The initial phase of a PE 3.0 mission is exploratory and only involves simple, low-risk tasks, such as orbital imaging or static landing. 
This reveals basic environmental constraints of the exploration target, such as the terrain properties, and narrows down the scientific questions to be investigated in the later phases of the mission. 
In the following phase, the space system adapts to the observed environment and engages in more complex tasks, such as surface and mobility, drilling and sampling, and even subsurface mobility, to perform deeper and more specialized, hypothesis-driven observations. While these phases themselves are similar to the missions included in a PE 2.0 approach, the key difference is that the adaptability is built into a single mission itself, instead of occurring on the ground between missions.

\begin{figure}
    \centering
    \includegraphics[width=0.5\textwidth]{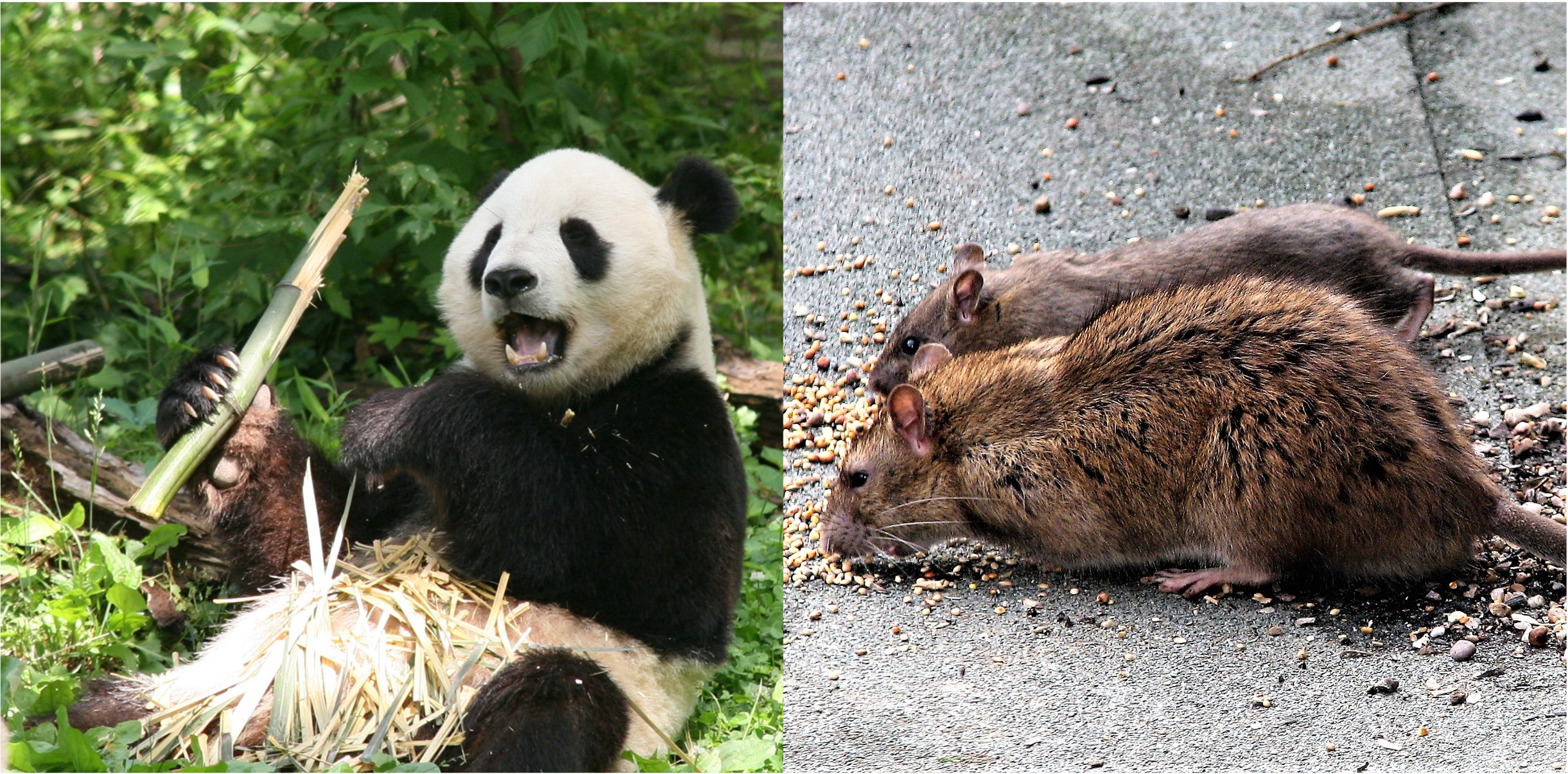}
    \caption{PE~2.0 is comparable to highly specialized animals such as pandas, while PE~3.0 envisions adaptive space systems like rats. Credits: Cliff/Patrick Roper. }
\end{figure}

An analogy can be drawn from biology.
PE~2.0 is like highly specialized animals, such as pandas, that depend on a singular or a small variety of diet and are very selective about environmental conditions. 
These animal species tend to be vulnerable to rapid environmental changes caused by humans. 
Some other animals are generalists, capable of digesting almost any food and living in a wide range of environments. 
For example, \textit{rattus norvegicus}, better known as brown rats or common rats, originally lived humbly in the grassy savannah in northern China and Mongolia.
They adapted to human-induced environmental changes, became commensal with humans, and eventually spread to every corner of the world except Antarctica. Although rats may not be the most beloved animals, PE~3.0 should be more like them than pandas.

Existing orbiter-lander missions and mission concepts, such as Viking, Hayabusa, OSIRIS-Rex, and Enceladus Orbilander \cite{mackenzie2021enceladus}, can be viewed as early examples of the proposed one-shot paradigm. 
However, the vision of PE 3.0 presented in this paper goes beyond them in terms of the approach and goals. 
Our approach is built-in adaptivity of space systems, as detailed in~\Cref{sec:system}, as opposed to opportunistic adaptation that has been common in past missions \cite{Ono2026}. 
In terms of the goals, our vision does not stop at static landing; rather, it includes capabilities like mobility, drilling, and sampling, which have been proven critical for answering some of the most important scientific questions such as habitability and existence of life, by past and ongoing Mars missions. 
For some planetary bodies, the main scientific target lies below the surface, most notably the icy moons with subsurface oceans.
In case of Enceladus, where ocean water is transported to orbital altitudes by geysers, it might be possible to detect biosignatures through orbital or surface sampling of the plume materials, which would prove the existence of alien life. 
A harder case to make would be inexistence. If biosignatures are not detected in the plume materials, it will require to go to the subsurface ocean to prove or disprove the existence. 
Furthermore, a positive detection of alien life could hardly be the end of our scientific quest. 
The next, substantially harder task is the characterization of life. 
The ultimate goal of PE 3.0 is the exploration of extremely remote targets, such as the Uranian and Neptunian systems, Kuiper Belt Objects (KBOs), interstellar objects, the Oort cloud, and even exoplanets. 
Unlike Mars, asteroids, and Jovian and Saturnian system objects, these targets have little to no existing proximity observations, and the information available through telescopic observations is extremely limited for planning space missions. 
Moreover, the distance to these objects alone precludes incremental PE 2.0-style exploration. 
Therefore, PE 3.0 would not be an option but a prerequisite if humanity wants to extend the reach of exploration to these distant destinations. 

\subsection{Paradigm Changes}
The transition from PE~2.0 to PE~3.0 will require multiple paradigm shifts and numerous technical innovations.  
This section presents three overarching themes that constitute the core concepts of PE~3.0: 1) transition from hypothesis-driven to capability-driven missions, 2) transition from hardware-defined systems to software-defined systems, and 3) transition from requirement-based systems engineering to less rigid processes driven by affordances and design principles. 

\subsubsection{From hypothesis-driven to capability-driven missions}
\label{sec:hypothesis-driven}
The separation of early exploratory and later hypothesis-driven missions in PE~2.0 must be dissolved in PE~3.0. 
The initial scientific goals of a typical 3.0 mission would be as open-ended as the first planetary missions such as Mariners, Pioneers, and Voyagers. 
However, rather than sending follow-on hypothesis-driven missions, in a PE~3.0 mission, hypotheses are dynamically formed and tested onboard the spacecraft as it learns about the environment. 

This means that the ongoing practice of defining a mission through a static scientific traceability matrix (STM) is not applicable to PE~3.0. 
An STM is a systems engineering artifact that explicitly links high-level science goals to measurement objectives, instrument requirements, and mission design, ensuring end-to-end traceability and verification of scientific return \cite{weiss2005science}. The STM emerged as a formalized framework in the 1990s alongside the maturation of NASA’s project management and systems engineering standards. In particular, the introduction of structured procedural requirements  \cite{NASA_NPR_7120_5, NASA_NPR_7123_1} institutionalized rigorous requirements traceability across mission lifecycles. Since then, STMs have become a standard and often mandatory component of NASA mission proposals and reviews, particularly within competed programs. 
However, the STM framework assumes that science objectives, environmental conditions, and measurement strategies can be defined and validated \textit{a priori}, resulting in a largely fixed and deterministic mapping from science goals to system implementation. This rigidity limits its applicability to adaptive missions operating in highly uncertain environments, where scientific priorities and measurement strategies may need to evolve dynamically in response to \textit{in situ} discoveries. \Cref{sec:science-definition} discusses more dynamic alternatives to the static STM-based PE~2.0 mission definitions and operations. 

In PE~2.0 missions, the design of onboard capabilities is driven by the scientific hypotheses through the STM.
In PE~3.0 missions, this is no longer the case. 
The scientific instruments are not selected for particular observations, but rather are chosen to support wide-ranging purposes, much like a Swiss army knife. 
For this reason, in PE~3.0, it makes more sense to standardize the spacecraft components and instruments rather than designing ones for every particular mission, and such standardization would perhaps help reduce the per-mission cost. 
Upon arrival at the destination, spacecraft activities and observations are planned to maximize the scientific gain within the available onboard capabilities and resources.
Adaptivity helps to expand the capability envelope with a limited set of instruments and components.

The contrast between PE~2.0 and PE~3.0 is somewhat similar to the difference between a business trip and a backpacking trip. A typical business trip has a pre-fixed schedule---you know exactly where to stay, who to meet, and what to do. You pack accordingly, bringing only what you know you will use (and somehow still forget your charger). In contrast, on a backpacking trip, you do not exactly know what will happen. (If you did, it would be a rather boring backpacking trip.) So instead, you pack generic, versatile items like ropes, duct tape, and a Swiss Army knife. You may not know exactly what you will use them for, but experience tells you they will come in handy—often in ways you could not have planned for.

\subsubsection{From hardware-defined to software-defined systems}

PE~3.0 requires radically adaptive space systems. We argue that such systems must be \textit{software-defined}. In PE~2.0, most spacecraft functions are hardware-defined and therefore not adaptable. For example, conventional antennas such as dipoles or parabolic antennas are hardware-defined because their radiation patterns are defined by their physical shapes. Adapting hardware-defined systems is difficult because it requires changing hardware configurations---extremely challenging, if not impossible, once a spacecraft is launched. In contrast, an active phased array antenna (APA) is a software-defined system since it can change its radiation pattern and even form multiple beams only through software commands. Such software-defined systems are much more adaptive in space because changing or augmenting their functions only requires commands or software updates. 

What makes a system software-defined is the hardware-level flexibility, reconfigurability, and multi-functionality that are controllable through software. For example, APA consists of an array of many antenna elements, each of which has its own transmit/receive module that can be independently controlled.  
Another example is field-programmable gate arrays (FPGAs). As opposed to conventional integrated circuits where circuits are baked in and fixed, FPGAs consist of a grid-connected array of programmable logic blocks that can be reconfigured through hardware description languages.

Some software-defined capabilities at the subsystem level are already in use in space, including FPGAs~\cite{Rieber2022} and software-defined radios \cite{reinhart2014space}. 
Modern science instruments are becoming increasingly adaptive, such as the MAss Spectrometer for Planetary EXploration/Europa (MASPEX) on Europa Clipper, which can tune the mass range(s) and resolution(s) of interest through fine control of its high-voltage ion optics \cite{waite2024maspex}. 
There are many other technologies at low technology readiness levels (TRLs) that are actively investigated and can be applied to future PE~3.0 missions, such as adaptable Hall effect thrusters, adaptable thermal management, and programmable metasurfaces. We will present representative samples of these technologies in~\Cref{sec:sw-defined-techs}.

It is often harder to make physical interactions with external environments software-defined, such as mobility, manipulation, and sampling. 
A viable approach is to incorporate redundant degrees of freedom (DOFs) in the system, which allows multiple mobility and manipulation modalities. 
Multi-agent systems can also achieve software-defined adaptivity by reconfiguring the physical formation and logical interactions between the agents.  
More radical approaches include modularized system and in-space self-manufacturing. These topics will be discussed in Sections \ref{sec:modularity}-\ref{sec:self-manufacturing}.

The shift to software-defined space systems also necessitates a corresponding advancement in onboard autonomy. As system flexibility and reconfigurability increase, the space of possible system states and behaviors expands dramatically, quickly exceeding what can be practically managed through traditional ground-in-the-loop operations. This challenge is compounded by communication constraints, including long latencies, limited bandwidth, and intermittent connectivity. As a result, relying on ground-based commanding to fully exploit the capabilities of a highly flexible system becomes increasingly infeasible. Instead, PE~3.0 requires onboard intelligence capable of making timely decisions about system configuration, resource allocation, and scientific operations in response to \textit{in situ} observations. \Cref{sec:intelligence} discusses this need in detail, including approaches to autonomous science, navigation, control, and embodied AI, which together can enable spacecraft to effectively utilize their software-defined capabilities in uncertain and evolving environments.

\subsubsection{Departure from requirement-based systems engineering}

In PE~2.0, space systems are designed to be ``just enough'' to satisfy a prescribed set of requirements, derived from partial knowledge of the target environment and bounded by pre-defined mission objectives. While this requirement-based paradigm provides rigor, traceability, and verifiability, it inherently assumes that system behavior can be fully specified a priori. 
This assumption breaks down in one-shot missions operating under deep uncertainty, where both environmental conditions and scientific priorities may evolve during the mission~\cite{Ono2026}. In PE~3.0, requirements naturally become dynamic, conditional, or even undefined at formulation, challenging the fundamental premise of requirement-based systems engineering. 

A key implication is that requirements can no longer serve as the sole organizing principle for system design. Instead, PE 3.0 calls for complementary or alternative paradigms that emphasize capability and adaptability over compliance.
Affordance-based design provides one such framework, in which system elements are characterized not only by their intended functions but by the range of possible functions they can support in unforeseen contexts~\cite{Maier2009}. This perspective enables systematic reasoning about in-flight repurposing and supports a shift from top-down, requirement-driven design to bottom-up, capability-driven architecting. Similarly, design approaches such as set-based design, real options, and catalog-based (over-)design prioritize deferred commitment, latent capability, and optionality, allowing the system to adapt its operational roles as uncertainty resolves.
\Cref{sec:design} will discuss these ideas in depth.

The departure from requirement-based thinking also extends to verification and validation (V\&V). Traditional verification focuses on demonstrating compliance with predefined requirements, an approach that is ill-suited for systems whose objectives and behaviors are not fully defined and/or expected to change. For adaptive systems, validation---i.e., demonstrating that the system achieves meaningful outcomes across a wide range of scenarios—becomes the dominant paradigm. 
This shift reflects the recognition that adaptability itself is not directly verifiable through discrete tests, but must instead be assessed through scenario-based, probabilistic, and often continuous evaluation approaches. 
\Cref{sec:vv} will elaborate alternative V\&V approaches for PE~3.0. 

Importantly, this departure does not imply abandoning rigor, but rather redefining it. In~PE 3.0, rigor is achieved not through exhaustive specification and verification of all anticipated behaviors, but through careful design of adaptable architectures, principled reasoning about uncertainty, and targeted validation of system capabilities under representative conditions. Requirements remain valuable where they can be meaningfully defined, but they are complemented by higher-level constructs—such as adaptability metrics, scenario-based specifications, and architectural design principles—that better capture the needs of radically adaptive, one-shot exploration missions.

\subsection{Outline of the Paper}
This paper is an outcome of a workshop hosted by the Keck Institute of Space Studies at the California Institute of Technology. 
The remainder of this paper is organized as follows. \Cref{sec:system} presents the systems engineering foundations of PE 3.0, including adaptive approaches to science definition and operations, dynamic and scenario-based requirements, design methods and principles for adaptive spacecraft, and challenges and approaches to verification and validation of adaptive systems. \Cref{sec:sdss} surveys enabling technologies for software-defined space systems, including reconfigurable hardware, multi-functional components, and modular architectures. \Cref{sec:intelligence} discusses the role of autonomy in PE 3.0, including onboard intelligence for science, navigation, control, and embodied AI. \Cref{sec:missions} then introduces representative mission concepts that illustrate the PE 3.0 paradigm, including a Neptune/Triton smart flyby, an ocean world explorer, and an Oort cloud reconnaissance mission. Finally, \Cref{sec:conclusion} concludes the paper by presenting key challenges, open questions, and future directions toward realizing radically adaptive planetary exploration.

\section{PE 3.0 Systems Engineering}\label{sec:system}
The following paragraphs discuss specific challenges related to how current SE processes can handle adaptive one-shot missions, focusing on science definition, science operations, requirements definition, design methods, and verification and validation. 

\subsection{Science Definition and Operations}\label{sec:science-definition}
\subsubsection{Science Definition}
In the current mission formulation paradigm, the Science Traceability Matrix (STM) for a mission concept starts from science goals and derives specific, quantified scientific objectives, from which measurement, instrument performance, and mission requirements are derived. The quantitative flow-down ensures that the science objectives can be achieved. However, in one-shot adaptive missions to the outer Solar System, the lack of \textit{a priori} knowledge of at least some science objectives---or their dependence on observations unavailable at concept development---is a major impediment to developing a standard STM. Even during the formulation of standard mission concepts, not all science objectives that will be accomplished by the mission necessarily appear in the STM. This is expected to be exacerbated for one-shot missions.

Nevertheless, even for one-shot missions, the present state of knowledge should be advanced enough to support a standard STM approach for at least a portion of science objectives. This standard approach should be followed to the extent possible, as it will give confidence that science objectives that can be adequately quantified can be met despite environmental uncertainty.

Given the challenges of incomplete science objectives, the traditional STM must be augmented. The traditional STM is quite effective when science questions are clear and lead directly to testable hypotheses. However, there are many situations where we do not have sufficient data to fully describe all STM elements until some mission data are returned. A situation may be: ``We think there is feature X in location Y, but we are not sure. If feature X exists, we would do measurement Z. Otherwise, we would do something else instead.'' A traditional STM is not robust to this uncertainty; a decision tree or conditional plan could potentially be a suitable alternative.

A prior example was provided by the Cassini mission: before launch, there were strong suspicions that Enceladus had a plume, but it was uncertain whether these jets existed, were continuous or transient, isolated or dense. Upon arriving and observing the plume, human operators decided to fly through the plume at multiple distances. The STM could have been more robustly formulated a priori as a conditional plan.

Conditional plans are more robust than rigid STMs, but they are still not enough, as it is generally not feasible to define a priori all possible cases. It follows that the STM will necessarily have to be updated by operators or by the spacecraft itself during the mission as discussed in the next section.  

\subsubsection{Science Operations}
Designing a science plan that is more robust to known uncertainty than the traditional STM is challenging; defining a concept of operations (ConOps) in light of such uncertainty may be even more difficult. 

For certain destinations, decisions  must be made too quickly for human-in-the-loop operations, necessitating autonomous science capabilities. Consider an interstellar probe which would have to survey a new Solar System upon arrival, make initial observations, and decide on further measurements without Earth-in-the-loop. Another case could be a vent and subsurface ocean probe at Enceladus or Europa~\cite{Chodas2023}, where science measurements may need to adapt based on the environment encountered---e.g., vent diameter being significantly smaller or plume dynamic pressure much larger than expected.


In these cases, it is important to hold much larger margin in resources and schedule to enable the desired flexibility and adaptability of the later mission ConOps to explore the most interesting hypothesis-driven science based on the data from the exploration phases of the mission.

ConOps strategies for adaptive missions must plan for and incorporate ways to replan and redirect efforts in light of unexpected data, environments, discoveries, and other uncertainties. Taking science observations in this manner will necessarily require prioritization of sampling and data collection. In particular, missions with high communications costs or low data rates will also need to be efficient about returning science data to Earth. Anomaly detection, such as AEGIS on the Curiosity and Perseverance Mars rovers~\cite{Verma2023}, is one method for determining which samples to take, as frequently the most interesting observations are the outliers, i.e., those that stand out from the surrounding environment. 

Once the interesting observations have been identified, some level of onboard processing of images or other types of data will be necessary. For example, cloud masks are used in Earth observation missions to avoid downlinking useless images, and image classification techniques are being explored for real-time detection of other Earth events such as floods, fires, or algal blooms \cite{Gorr2024ReactiveWebs,Nag2020D-SHIELD:Decisions} or detection of neighboring spacecraft~\cite{grauer2024visionbaseddetectionuncooperativetargets}. Data compression algorithms, including lossless and lossy techniques, can also be used onboard to reduce the size of images. 

In addition to onboard processing of the data,  strategies and protocols must be in place for prioritizing which data are downlinked, potentially providing onboard adaptability within the prioritization process itself. An adaptive system would allow for making opportunistic measurements without a human in the loop (e.g., collecting images of a dynamic feature such as a fresh impact, or augmenting a flyby trajectory to aim for a collimated jet of a planetary plume). In addition, certain bodies may evolve over time, perhaps as a function of season or orbital period, and so making multiple measurements in the same place but at different times may yield important discoveries. Typically these are established early in the mission STM and broken out by phase, but new observations or evolved knowledge of the system being studied (either via the spacecraft itself or perhaps new experimental or modeling results on the ground) may lead to reprioritization, which an adaptive prioritization system could do autonomously. This kind of autonomous task planning and scheduling by spacecraft and spacecraft swarms is becoming a mature technology thanks to ongoing research and development activities and several onboard deployments \cite{Candela2023DynamicMissions,Gorr2024ReactiveWebs,zilberstein2025federated,aguilar2025decentralized,David20223D-CHESS:Systems,herrmann2023reinforcement,doi:10.1126/scirobotics.adi3099,doi:10.2514/6.2024-4889}. More on these technologies is discussed in \Cref{sec:sw-defined-techs}. 

Finally, extraordinary claims require extraordinary evidence, and any ConOps aiming to verify extraordinary claims must account for this requirement. Backing such extreme and/or low-probability observations with more supporting evidence will also lead to more data being returned to Earth. Investing in development of ``AI scientists'' pre-launch will enable on-the-fly determination of which additional data are required to be captured and returned to substantiate extraordinary claims. Ensuring that the returned data and conclusions are trusted will require involvement from broad science teams during the pre-flight learning phases of the AI to establish trust in the onboard AI to do the right work during operations. Ultimately, the goal is for the onboard AI to evolve beyond heavy reliance on the ground team, much like a PhD student who initially depends on peers and advisors for guidance, but gradually develops into an independent researcher capable of verifying their own claims with high confidence and provable reliability.

\subsection{Requirements}
\label{sec:requirements}
Adaptive missions operate under persistent uncertainty, requiring them to accommodate change rather than assume stability. Requirements in adaptive space missions may evolve for both intrinsic and extrinsic reasons. Intrinsic changes arise from failures, degradation, or improved understanding of onboard behavior discovered during operations, while extrinsic changes may be driven by environmental uncertainty, evolving operational context, new mission objectives derived from operational insight, new capabilities, or new scientific discoveries. These sources of change span different levels of uncertainty, from known unknowns to unknown unknowns, and may therefore require different problem formulations and corresponding verification and validation approaches. They also span different time scales, which has implications for the selection of the relevant solutions. 

Using a requirements-based systems engineering approach in this context leads naturally to requirements that are dynamic, conditional, or both. A central challenge is therefore how such requirements can be formulated, managed, and assessed over the mission lifecycle. The development of operational concepts and scenarios becomes particularly important in grounding requirements in an operational context. Evaluating scenarios together with system characteristics provides a concrete means to characterize capability under changing conditions. Scenarios that explicitly capture interactions between subsystems and the need for adaptability are especially valuable in making dynamic requirements operationally meaningful.

Two complementary paradigms can be used to define adaptability requirements. The first relies on semantics associated with change scenarios \cite{ross2015towards,ross2019ilities}. In this approach, adaptability is specified through a set of representative scenarios composed of a change impetus, the current system state, and the desired end state or behavior that the system must exhibit. While this does not guarantee adaptability to all unforeseen situations, the assumption is that a system capable of adapting to a sufficiently rich set of scenarios is likely to adapt to related, previously unanticipated ones. From a value-based perspective, adaptability contributes to expected mission value, where the perceived likelihood and impact of future scenarios influence which cases are emphasized during design.

The second paradigm prescribes architectural or design principles that are understood to enable adaptability, or conversely avoid mechanisms known to inhibit it. These principles are typically associated with system “ilities,” such as robustness, resilience, and flexibility \cite{colombo2017classification}. Unlike traditional performance requirements, such properties are not directly verifiable through a single test or metric. Instead, they are realized through architectural patterns, structural choices, and degrees of freedom that facilitate certain behaviors under uncertainty. Verification in this context focuses on demonstrating that the relevant patterns are correctly implemented, that they do not conflict with one another, and that their presence is supported by evidence linking them to adaptability and other relevant "ilities". Importantly, principles are rarely binary; they facilitate adaptability to varying degrees rather than acting as simple enablers or inhibitors.

This perspective also motivates a shift from purpose-based to affordance-based modeling \cite{Maier2009,kannengiesser2012process}. Traditionally, system elements are characterized according to their intended purpose. For adaptive systems, it is important to characterize components in terms of what they can potentially be used for beyond their original function. Such affordance-based descriptions support systematic reasoning about repurposing during operations and enable reuse of models across mission phases. Fundamentally, an adaptive system is one that can dynamically allocate functions or purposes to different components in flight; the architecting process should therefore reflect this flexibility and use abstractions consistent with those employed during operations.

It is important to note that adaptability is not required uniformly across all systems or subsystems. It is typically justified where environmental uncertainty is high, and it comes at a cost in mass, power, computational complexity, and design effort. Adaptability can be linked to degrees of freedom within the system: additional actuation, configurability, or functional overlap can increase flexibility but also introduce complexity. Broad performance requirements across wide operating ranges may result in infeasible designs if not carefully assessed, making early feasibility checks particularly important for adaptable systems.

To support these adaptability-cost-risk trade-offs in a structured manner, it is useful to introduce ways to characterize and communicate levels of adaptability within requirements, ranging from qualitative to quantitative. For example, a semi-quantitative System Adaptivity Level (SAL) could be introduced, analogous to Technology Readiness Level. SAL is conceived as an integer scale defined descriptively rather than through a single quantitative metric. It captures multiple dimensions of adaptability, including tolerance to environmental uncertainty, anomaly accommodation, flexibility in operational objectives, and repurposing capability. SAL may be defined at the system or subsystem level and incorporated directly into requirements, serving both as a design target and as an interface between stakeholders. While its precise definition remains future work, illustrative levels clarify intent: a low level (e.g., SAL-1) may correspond to a subsystem that tolerates up to a specified percentage change in key environmental parameters; an intermediate level (e.g., SAL-5) may accommodate qualitatively different operational objectives; and a high level (e.g., SAL-10) may support entirely new purposes beyond the original design intent. In this way, SAL provides a structured means to reason about adaptability as an “ility,” relate it to architectural choices, and connect requirement formulation to V\&V considerations.

\subsection{Design Methods for Adaptive Spacecraft}
\label{sec:design}
Traditional "point-based" design identifies fixed requirements and allocates them to specific hardware. To enable adaptivity in one-shot missions, we must shift toward methods that prioritize latent capabilities and deferred commitment. As illustrated in \Cref{fig:panda_vs_rat}, the standard design process in PE 2.0 (illustrated by the panda) is rigid and typically ends before the mission is launched, with the design being frozen then (with the possible exception of deferred software deliveries). Conversely, design in PE 3.0 missions (the rat) will never be strictly frozen and the design process never stops. Software-defined components can allow a continuous redesign process to adapt to new conditions. In-space autonomous assembly and manufacturing, perhaps combined with \textit{in situ} resource extraction and utilization, can even enable hardware redesign. 

\begin{figure}
    \centering
    \includegraphics[width=0.9\linewidth]{figures/design_paradigms.jpg}
    \caption{In PE 2.0, design is frozen before launch and the design process ends there. In PE 3.0, the boundaries between mission phases are blurred, where the design process is continuous and the design is never frozen.}
    \label{fig:panda_vs_rat}
\end{figure}

While new design methods are needed for PE 3.0, it is not necessary to start from scratch. There are several existing design methods that can be combined and adapted for designing adaptive one-shot missions. They are discussed in the following paragraphs. 

\textit{Affordance-Based Design}~\cite{Maier2009} focuses on the interactions between an agent and an object, identifying unintended but inherent functions of a component---providing a structured way to discover alternative uses for hardware already on board.
In essence, the design of adaptive one-shot missions necessitates a move from Top-Down (Requirements-Driven) to Bottom-Up (Capabilities-Driven) systems engineering. A camera is no longer just a ``science imager''; it is a ``source of optical data'' that can serve navigation, self-diagnostics, or opportunistic science.

A related approach is \textit{Set-Based Design}~\cite{Toche2020}. Rather than finalizing a single design early, Set-Based Design maintains a set of compatible design solutions, narrowing them only as uncertainty resolves. This deferred commitment preserves mission agility but requires rigorous V\&V to ensure all potential paths remain viable, making it very relevant for the design of adaptive one-shot missions. 

Designing adaptive one-shot missions is primarily about designing for flexibility and changeability under uncertainty. There is a rich literature in this area. \cite{DeNeufville2003} analyzes flexibility in design using the framework of real options in finance, where for example one can purchase an option (but not an obligation) to buy/sell stock at a given price; so one can decide whether or not to exercise the option when some of the uncertainty resolves. The analogy in the engineering world is to invest early in the development of additional assets (interfaces, infrastructure, capabilities) or capacity that will provide the option later to change the design. For example, building additional physical interfaces (e.g., spare ports) so that additional devices can be connected later if they are available. 

\textit{Epoch-era analysis} \cite{ross200811} is another related framework in which a set of change scenarios (epochs) are defined and systems are evaluated in timelines defining sequences across epochs, called eras. 
Common to all these methods is the idea of defining a set of scenarios and designing a system to maximize some uncertain metric such as probability of success, which will generally require bringing more capability than strictly necessary given the current knowledge.       
\textit{Catalog-Based (Over-)design} is often used in spacecraft design already as a measure to reduce cost and risk. Selecting from standardized catalogs often results in excess performance or functionality not strictly required by the primary mission. In adaptive contexts, this overdesign is an asset, providing the resource margins (power, data, mass) necessary for reconfiguration.

The concept of modularity can be applied during the development phase to leverage the cost and risk reductions through commonality, reuse, and product platforms ~\cite{Simpson2001}. This was popularized in the automotive industry but has also been adopted in the aerospace industry. The DARPA F6 program, in which a classical spacecraft was decomposed into a set of free-flying heterogeneous spacecraft, used commonality and spacecraft subsystem platforms to reduce the cost and risk of the concept \cite{Brown2009Value-CentricProgram}.

While commonality can be used to reduce the cost of adaptive one-shot missions, another interpretation of modularity, namely modularity for operational reconfigurability, is even more directly relevant. Siddiqi and De Weck reviewed some modeling methods and design principles for reconfigurable systems \cite{siddiqi2008modeling}. Several design methods have been developed for reconfigurability in the sense of changing orbit or trajectories, e.g., to optimize coverage performance of a constellation under spacecraft failures or adversarial attacks. But reconfiguration spans more than just trajectory changes, and can affect any tunable parameter of any spacecraft component, including instruments, radios, etc. Reconfigurability could potentially also be achieved at the platform level. More discussion on reconfigurability technologies is provided in \Cref{sec:modularity}.  

\subsection{Design Principles}
As proposed earlier in \Cref{sec:requirements}, PE 3.0 may need to rely less on rigid requirements and more on qualitative design principles. For the purposes of this paper, we define a design principle as a piece of prescriptive design guidance that is likely to lead to an improved probability of mission success. This definition is related to the one provided in \cite{Fu2016DesignDirections} and related concepts such as design heuristics \cite{fillingim2020design} or design patterns \cite{alexander1977pattern,gamma1993design}. Design principles are useful because they are reusable pieces of knowledge that generalize across missions and projects. Aerospace organizations like NASA's Jet Propulsion Laboratory (JPL) and Goddard Space Flight Center have processes and tools in place to maintain and leverage databases of design principles.

In our KISS workshop, several design principles for adaptive one-shot missions were derived, based on a combination of existing system architecture principles from the literature \cite{maier2009art,crawley2015system,10.1007/978-3-031-51928-4_22} and expert judgment. Some of these principles were already laid out in a previous publication by some of the authors \cite{Ono2026}. Brief descriptions of the design principles are provided here.


\paragraph{Principle of Extra Sensing Modes.}
Adaptive space systems should include redundant and diverse sensing modalities beyond immediate mission requirements. Additional sensing enables the diagnosis of anomalies and adaptation to unexpected environments by providing alternative sources of information when primary sensors fail or are insufficient. This applies to mission and science sensors, guidance, navigation, and control (GNC) sensors, and telemetry systems that can help diagnose failures. However, incorporating additional sensing increases system complexity, mass, power consumption, and cost, and imposes additional burdens on data handling, integration, and verification. Historical examples illustrate the value of this principle: Galileo used its Sun Gate sensors to diagnose antenna deployment issues; Deep Space 1 repurposed a science camera as a star tracker; OSIRIS-REx relied on backup navigation cameras to overcome LiDAR limitations; and InSight’s limited sensing capability hindered its ability to adapt during mole deployment.

\paragraph{Principle of Redundancy in Mechanical and Electrical Degrees of Freedom.}
Adaptive space systems should incorporate additional mechanical and electrical degrees of freedom beyond those strictly required for nominal operation. These additional degrees of freedom enable alternative modes of operation and improvised solutions when components fail or new requirements emerge. This principle applies to mechanisms such as robotic arms and actuators, as well as to electrical architectures including power routing and switching. The trade-offs include increased mass, system complexity, and potential failure points, as well as a higher V\&V burden. This principle may conflict with traditional design philosophies that seek to minimize moving parts. Nevertheless, past missions demonstrate its utility: InSight used a scoop that was included in its inherited design (despite not being required for the nominal mission) to assist with mole recovery; Hayabusa leveraged independent power supplies to cross-wire thrusters; SMAP used redundant switches in recovery attempts; and Galileo’s lack of a reversal mechanism limited recovery options after the loss of its high gain antenna.

\paragraph{Principle of Component and Subsystem Interconnectivity.}
Adaptive space systems should enable flexible data, electrical, and functional interconnections between subsystems beyond strictly necessary interfaces. Such interconnectivity allows subsystems to be repurposed and combined in novel ways to address unforeseen challenges or opportunities. This principle is relevant to both hardware and software interfaces, particularly in systems where subsystems may need to exchange data or functionality not originally anticipated. The main trade-offs include increased integration complexity, the potential for unintended interactions or faults, and reduced subsystem independence and modularity. Despite these challenges, interconnectivity has proven valuable in practice: Hayabusa cross-wired ion sources and neutralizers across thrusters; Juno’s star tracker provided raw images for scientific use; and Perseverance transferred data between onboard computers to enable new capabilities.

\paragraph{Principle of Additional Resource Margins.}
Adaptive space systems should include margins in system resources—such as fuel, power, bandwidth, memory, and computational capacity—beyond what the minimums dictated by standard best practices. These additional margins provide the capacity needed to implement adaptations, recover from anomalies, and extend mission capabilities. This principle applies to both physical resources and digital resources. The trade-offs include increased mass, cost, and design constraints, as well as potential inefficiencies if the margins are never utilized. However, many missions have benefited from such margins: Hayabusa’s extra xenon enabled cross-operation of thrusters; extended missions such as Galileo, Juno, and Deep Space 1 relied on fuel reserves; limited onboard computing constrained Galileo and required the use of auxiliary processors; and communication bandwidth has constrained the deployment of new capabilities on Perseverance.

\paragraph{Principle of Onboard Software Reconfigurability.}
Flight software should be designed to allow modification of most parameters through command interfaces during operations. Functional redundancy and resource margins cannot be effectively exploited without corresponding flexibility in the software that governs system behavior. This principle applies to command interfaces, parameter tuning, and access to low-level system functionality during the mission. The trade-offs include increased risk of operator error, higher verification complexity, and potential safety concerns, which necessitate strict operational procedures and testing. In practice, most missions already rely on some degree of software reconfigurability, whether through low-level command access, tunable parameters, or, in some cases, direct interaction with onboard computing environments.

\paragraph{Principle of Reprogrammability of Onboard Computers.}
Space systems should support in-flight reprogramming of onboard computers to enable the introduction of new capabilities and major behavioral changes not anticipated during design. This capability extends to both primary and auxiliary processors, including those used for imaging and attitude determination and control. While reprogrammability enables significant adaptability, it introduces risks if updates affect critical systems and requires robust validation, redundancy, and rollback mechanisms, as well as sufficient compute and communication resources. Historical examples highlight its importance: Galileo implemented onboard data compression through software updates; Deep Space 1 reprogrammed its software to use the MICAS instrument as a star tracker; and Perseverance leveraged an auxiliary processor to enable new capabilities, with redundant computers helping mitigate update risks.

\paragraph{Principle of Capability Preservation (Non-Preclusion).}
System-level design should preserve the full intrinsic capabilities of components and avoid constraining or disabling latent functionality. Many adaptive opportunities rely on exploiting capabilities that were not originally intended for the baseline mission, and precluding them at the system level eliminates these opportunities before launch. This principle primarily applies to system architecture decisions, including electrical interfaces, thermal limits, and software drivers, which may unintentionally restrict component performance envelopes. The trade-offs include the need for additional design margins in power, thermal, and data handling resources, increased integration complexity, and potentially higher cost due to less tightly optimized designs. However, past missions demonstrate the value of preserving latent capabilities: Juno’s star tracker could be used for science because access to raw data was preserved, and Galileo leveraged processors and subsystems beyond their originally intended roles.

\paragraph{Principle of Standardization and Multi-functionality.}
Adaptive space systems should employ standardized, modular components with multi-functional capabilities to maximize utility per unit mass. Standardization enables reuse and flexibility across mission phases, while multi-functionality reduces the need for mission-specific hardware and increases the likelihood that components can be repurposed during anomalies or mission extensions. This principle applies broadly to instrument and spacecraft component selection. The trade-offs include potential performance compromises relative to highly specialized components, increased software and integration complexity, and the risk of overloading components with competing functions. Nonetheless, numerous missions have demonstrated that adaptations often rely on components that are not tightly specialized, with cameras, processors, and other subsystems frequently repurposed for roles beyond their original intent.

\paragraph{Principle of Hypothesis-Free Instrumentation.}
Missions should include a diverse and complementary set of instruments not tied to a single scientific hypothesis, enabling broad environmental characterization. In highly uncertain environments, pre-defined hypotheses may be incomplete or incorrect, and broader sensing capabilities enable discovery while supporting adaptation to unexpected conditions. This principle is particularly relevant to remote and \textit{in situ} science instrumentation for the exploration of poorly characterized environments, such as ocean worlds, gas giant planets, or the Oort cloud. The trade-offs include increased payload mass and power requirements, the difficulty of selecting instruments without some guiding hypotheses, and a greater burden on data processing and analysis. While fully hypothesis-free designs are challenging in practice, missions that incorporate diverse and complementary sensing modalities are generally better positioned to adapt to unanticipated findings and evolving scientific objectives.

This list of principles is work in progress and will be refined in a subsequent workshop. 

\subsection{Verification and Validation}
\label{sec:vv}
Formulating adaptability requirements raises a central question: how should V\&V be structured when system behavior may expand or evolve under uncertainty? For adaptive systems operating under high uncertainty, exhaustive testing of all possible behaviors and conditions is infeasible due to cost, schedule constraints, and the combinatorial growth of system states and subsystem interactions. This challenge is not entirely new. In previous planetary missions, autonomous fault response cases have been too varied and complex to test exhaustively, and complex subsystems such as rover navigation already involve parameter combinations that exceed what can be individually evaluated. In such cases, engineering teams rely on representative and/or bounding test cases to establish confidence without implying complete coverage.

Adaptive systems further amplify this challenge. As autonomy increases, system behavior depends on sequences of observations and decisions made during operations, leading to a proliferation of possible states and interactions. At the same time, increased processing capability and the need to generate training data for AI-enabled autonomy allow far larger numbers of mission-level simulations to be executed. In this sense, the V\&V of adaptive space systems can be viewed as an extension of existing methodologies, but at a much larger scale: while more scenarios can be explored computationally, the resulting volume of simulation output quickly exceeds what engineers can realistically review manually. 

However, V\&V need not remain confined to demonstrating compliance with predefined behaviors; it can be reconceptualized as the assessment of system capabilities. In this view, the emphasis shifts from confirming that a system produces specific outputs under specified conditions to establishing that it possesses the capacity to do something (i.e., adapt). An illustrative analogy is learning. Traditional V\&V approaches resemble examinations of retained knowledge; that is, evaluating whether a system can reproduce expected responses. In contrast, a capability-oriented perspective seeks to verify the system’s ability to acquire and generalize knowledge.

In the context of adaptable systems, validation becomes more significant than verification: verification ensures that the system meets its specified requirements, while validation assesses whether the system behaves appropriately in its intended operational context. For adaptive systems, where behavior emerges from interactions and may vary across scenarios, demonstrating correct system-level behavior becomes more critical than verifying compliance with predefined requirements alone. An analogy can be drawn with how human operators, such as pilots, are validated: confidence is built by demonstrating acceptable behavior within defined operational bounds rather than by testing every possible situation. This perspective may likewise inform how confidence is established in autonomous systems. Because system-level behavior emerges from subsystem interactions, particular attention must therefore be given to potential undesirable couplings introduced by adaptability mechanisms, which must be understood and, where necessary, constrained.

The traditional “test-as-you-fly” principle must also be reconsidered. A “test-while-you-fly” approach may be appropriate for certain missions, in which some capabilities are validated incrementally as they become operationally relevant. This is feasible for slow-paced missions such as planetary rover surface operations, where extended timelines allow adjustment over time, but not for short, irreversible mission phases such as planetary flybys, where behavior must be correct at first execution. More generally, as mission goals and operational contexts evolve, validation must become continuous, recognizing that prior validation results may no longer fully characterize future behavior.

One possible solution to these limitations is to leverage AI to assist the V\&V process. As the behavioral space expands beyond what can be reasonably anticipated through human-designed test campaigns, and simulation campaigns generate more data than can be reviewed in detail, intelligent tools may be required to support scenario generation and scalable processing and analysis of simulation results. For scenario discovery, adversarial methods such as Adaptive Stress Testing ~\cite{Koren2019} use intelligent search to steer closed-loop system simulations toward rare or safety-critical conditions. Related adversarial and simulation-based frameworks are being explored within the research community and at NASA centers, including JPL, to support stress testing of highly autonomous systems. Recent work includes lifecycle assurance processes for autonomous space systems ~\cite{Pinto2026} and simulation-based adversarial testing approaches for adaptive mission scenarios~\cite{Ethvignot2026}.

\section{Software-Defined Space Systems}
\label{sec:sdss}

Next, we turn to the technological enablers that make Planetary Exploration 3.0 (PE 3.0) feasible in practice. As discussed in Section I, one of the key requisites of PE 3.0 is to make space systems software-defined, enabling them to adapt their functions in response to uncertain and evolving environments. At its core, a software-defined system leverages hardware-level flexibility that can be exposed and reconfigured through software. There are multiple pathways to achieving this flexibility, among which reconfigurability (\Cref{sec:sw-defined-techs}) and multi-functionality \Cref{{sec:multi-functionality}} play central roles. These capabilities allow a single system to support a wide range of operational modes beyond those anticipated at design time. However, extending such adaptability to robotic systems that physically interact with \textit{in situ} environments—such as mobility, manipulation, and sampling—remains particularly challenging due to mechanical complexity and environmental uncertainty; these challenges and approaches are discussed in \Cref{sec:modularity}. Looking further ahead, the ultimate form of adaptivity may lie in systems capable of in-space self-manufacturing and reconfiguration, enabling not only functional evolution but also physical transformation of the system itself \Cref{{sec:self-manufacturing}}.


\subsection{Reconfigurable Space Hardware Technologies}
\label{sec:sw-defined-techs}
There are many reconfigurable hardware technologies for space applications with widely varying technology readiness levels. Presented in this subsection are the samples that are representative, but not exhaustive, of the relevant technologies. 

\textbf{Software-defined Radio}: Traditional radio made of analog circuits has very limited adaptability. For example, a frequency-modulateion (FM) radio cannot turn to amplitude or phase modulations without replacing the circuit. Software-defined radio replaces the analog circuits with programmable digital circuits by a varying degree, hence it can alter operating frequency, radiation pattern, or polarization through software, enabling the same hardware to support multiple mission functions. Some modern software-defined radio uses direct waveform sampling operating at multi-GHz sampling frequencies to fully digitalize the signal processing. Had Cassini had a software-defined radio, it could have solved the Doppler shift issue of Cassini-Huygens relay discovered after the launch mission. Many modern spacecraft are already equipped with software-defined radio. For example, Mars Reconaissance Orbiter, launched in 2005, doubled the data rate for relay communication with surface assets (rover and landers) after the launch owing to the flexibility of software-defined radio. 

\begin{figure}
    \centering
    \includegraphics[width=0.6\linewidth]{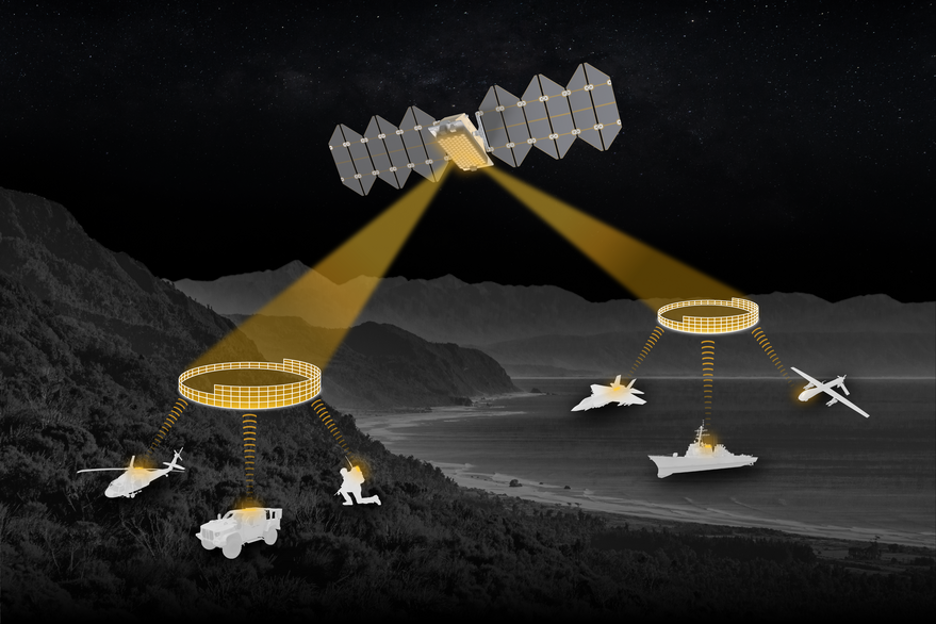}
    \caption{Active phase array technology allows forming multiple beams and communicating with multiple geologically distant nodes simultaneously. Image: CeciumAstro.}
    \label{fig:phased_array}
\end{figure}

\textbf{Phased-Array Antennas}: The most important defining factor of traditional antennas is physical shape. A parabolic antenna has a focused beam while a dipole has omnidirectional sensitivity. Modern phased-array antennas allow beam steering and shaping through electronic control rather than mechanical pointing, enabling spacecraft to dynamically reallocate communication capacity or retarget ground links without physical reconfiguration. Recently, active phased array (APA) technologies have emerged, in which each antenna element has an independent radio transceiver. APA allows more drastic adaptations, such as forming multiple beams concurrently (Figure \ref{fig:phased_array}). In 2021, CesiumAstro’s Cesium Mission 1 successfully tested the APA technology in space on two cubesats. The company markets in-space APA systems, as well as APA satellites.  

\textbf{Adaptable Thrusters}: The propulsion subsystem is perhaps among the hardest to make software-defined, but some technology exists. For example, there are two types of Hall thrusters: stationary plasma thruster (SPT) and thruster with anode layer (TAL). The difference is the material on the surface of the channel wall. SPT uses ceramic, which is non-conductive, while TAL uses conductive metal. The two types of Hall thrusters have complementary strengths and weaknesses. SPT has a lower specific impulse (i.e., lower fuel efficiency) but a higher thrust, while TAL has a higher specific impulse but lower thrust. Levchenko et. al. \cite{Levchenko2018} proposed a thruster with smart nanomaterials, which can change between SPT and TAL through hydrogeneration and dehydrogeneration of the coating material of the channel wall, hence allowing a PE 3.0 mission to optimally adjust the balance between specific impulse and thrust in different mission phases. 

\textbf{Adaptable Thermal Management}: Thermostats are essential building blocks for thermal control. Conventional spacecraft used hardware switches such as bimetallic switches, which are simple and reliable but had no ability to adjust the temperature setpoints. Modern spacecraft often use software switches that read from temperature sensors. Mars Helicopter Ingenuity successfully survived a winter even though it was not designed to do so because the ground operation lowered the temperature setpoint and saved energy. Thermal control of future spacecraft could be made more adaptive by new technologies that are currently under research, such as variable conductance heat pipes \cite{Kravets2017} and morphing space radiators \cite{Bertagne2018}.

\textbf{Computing and Avionics}: Most modern computer architectures are already general-purpose and readily reprogrammable, and spacecraft have taken advantage of this capability for decades. Since the widespread adoption of onboard digital flight computers in the 1980s and 1990s, spacecraft have routinely uploaded new software during missions to modify operational behavior, correct faults, and introduce new capabilities. Many missions have adapted to new capability needs by sending flight software updates from the ground, such as the addition of a data compression algorithm to Galileo as a mitigation for the HGA deployment failure and Deep Space 1’s repurposing of a science camera as a star tracker \cite{Ono2026}. Reprogrammable onboard computing also enables many of the adaptable sensing and communication systems described above, since changes in signal processing, instrument configuration, or operational logic can often be implemented entirely in software without altering the underlying hardware. A very recent example is NASA’s Perseverance rover, which repurposed the onboard computer originally used to operate the Ingenuity helicopter to instead run terrain-relative navigation software that allows the rover to autonomously determine its precise location on Mars after the helicopter mission ended.

\begin{figure}
    \centering
    \includegraphics[width=0.65\linewidth]{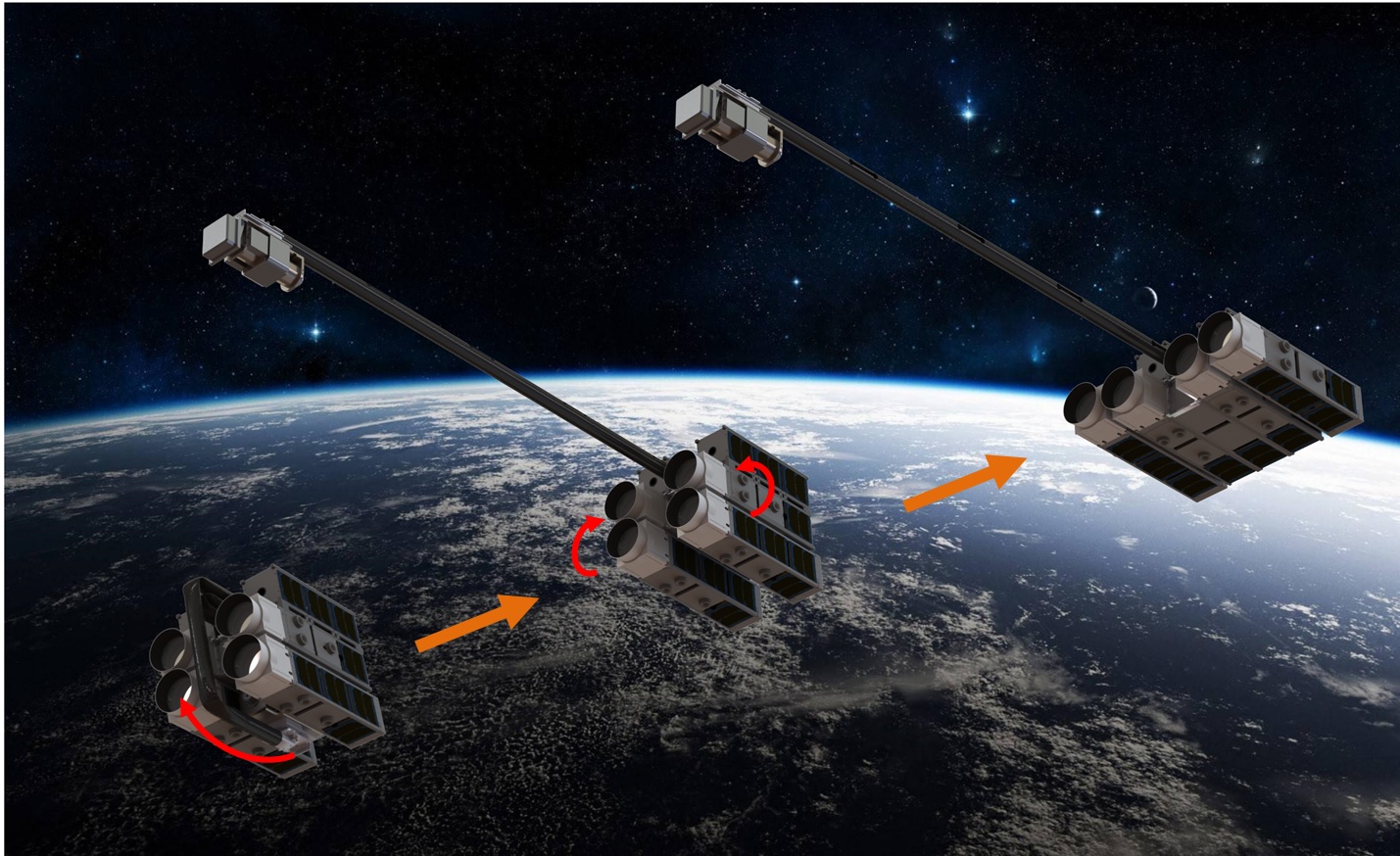}
    \caption{AAReST (Autonomous Assembly Reconfigurable Space Telescope) concept. \cite{Underwood2018}}
    \label{fig:aarest}
\end{figure}

\textbf{Remote Sensing}: Reconfigurable sensing systems provide a similar form of adaptability. Adaptive optics systems can adjust optical characteristics to compensate for disturbances or to optimize imaging performance under different observing conditions or spacecraft configurations. For example, the DeMi mission has demonstrated adaptive mirrors on a CubeSat platform \cite{Morgan2021}. Similarly, the AAReST technology demonstration mission concept (\Cref{fig:aarest}) proposed large scale shape adaptive optics for a reconfigurable space telescope \cite{Underwood2018}. Reconfigurable RF instruments can dynamically change observation modes, for example switching between communication and radar sensing functions using the same antenna hardware. Reconfiguration of operating frequency, polarization, and radiation pattern can be achieved using electronic switches, phase changing materials, or physical reconfiguration of conductors~\cite{Beneck2021}. Advances in tunable optical and RF components further extend this capability by enabling spacecraft instruments whose sensing characteristics such as wavelength range, field of view, or spatial resolution can be modified through electronic control. A prominent example is the Cassini RADAR instrument, which supported multiple operating modes including synthetic aperture radar imaging, altimetry, and scatterometry using the same antenna and hardware, with the instrument behavior determined largely through software configuration~\cite{Elachi2004}.

\textbf{Adaptable Optics}: Optical devices are used for engineering and scientific purposes in space, from attitude determination via star trackers, optical communication, and stereo perception for autonomous driving to scientific imaging and spectroscopy. While the basic working principle is common - i.e., collect and focus photons on a sensor array, they differ substantially in various characteristics such as focal length, wavelength, sensitivity, and dynamic range. Adaptive optics systems can adjust these optical characteristics to adapt their functionalities and optimize imaging performance under different observing conditions. Recently, metasurface technologies have emerged, which use 2D arrays of subwavelength-scale structures ("meta-atoms") designed to manipulate electromagnetic waves in unprecedented ways. Some metasurfaces are reconfigurable (e.g., \cite{Popescu2026}), capable of altering optical characteristics by applying heat or electrical voltages on meta-atoms, for example. Some reconfigurable metasurfaces can even modulate the polarization and form multiple beams \cite{Shirmanesh2020}. In space, metasurfaces can potentially replace rotating filter wheels, observe multiple directions simultanously, and enhance the data rate of optical communication. Furthermore, reconfigurable metasurfaces can possibly enable a single optical system to perform many different functions. A major drawback of existing metasurface technology is its very low efficiencies - typical reflectance is as low as 10\%.

\subsection{Multi-functionality}
\label{sec:multi-functionality}

Repurposing spacecraft subsystems is not new. Perhaps the most well-established example is the use of spacecraft telecommunications systems for radio science. The same radios and antennas used for communication with Earth can also be used as scientific instruments by analyzing Doppler shifts and signal distortions in the transmitted signal~\cite{deep_space_network_instrument}. The Huygens probe used its communications link to Cassini to perform a Doppler wind experiment, inferring wind speeds in Titan’s atmosphere by measuring small frequency shifts in the transmitted radio signal during descent~\cite{huygens_wind}. Repurposing also occurs within sensing systems. Cameras originally designed for spacecraft navigation or engineering monitoring are frequently used for scientific observation once operational tasks are complete. For example, the navigation and hazard-avoidance cameras on the Mars Exploration Rovers, Curiosity, and Perseverance were designed primarily to support rover mobility and landing operations, but they have also been widely used for geological imaging, atmospheric observations, and monitoring surface processes such as dust devils and cloud formation. Similar multifunctionality is common across planetary missions, where cameras and other engineering sensors used for navigation or targeting during cruise and landing phases are later repurposed to collect scientifically valuable observations. 

Another approach to multifunctionality is the integration of sensing, communication, or thermal functions into components that historically served only structural roles. This follows the principle of \textit{no wasted mass}; machines can be built more mass-efficient if their structural elements could be made to serve one or more secondary roles. A notable line of research in this direction explores structural batteries, which are load-bearing components that also serve as chemical energy storage devices~\cite {structural_batteries1,structural_batteries2}. By supporting spacecraft loads with structural batteries as opposed to pure-structure components, the spacecraft can reduce or completely remove reliance on single-purpose battery elements. Less research has been conducted on structural components that implement the functions of thermal protection systems~\cite{hotstructure}, structural components with embedded sensing for health monitoring~\cite{fibreoptic_sensing}, and structural components with many secondary roles like heating, sensing, actuation, circuitry, and radiation shielding~\cite{esa_multifunction}. 

\subsection{Software-defined Robotic Systems}\label{sec:modularity}

Robotic systems that directly interact with \textit{in situ} environments, such as surface and subsurface mobility, manipulation, and drilling, are particularly difficult to be reliably adaptive. 
One of the reasons is the mechanical complexity. 
Another underlying reason is the substantial uncertainty in terramechanics, or mechanical interaction with terrains, as represented by the examples of InSight's failed mole penetration and Perseverance's first coring attempt, as mentioned in \Cref{sec:intro} (Also see \Cref{fig:unknowns}-e,f). This section presents approaches to making \textit{in situ} robotic systems adaptive. 

\begin{figure}[t]
\centering
\begin{subfigure}[t]{0.51\textwidth}
\centering
\includegraphics[width=\linewidth]{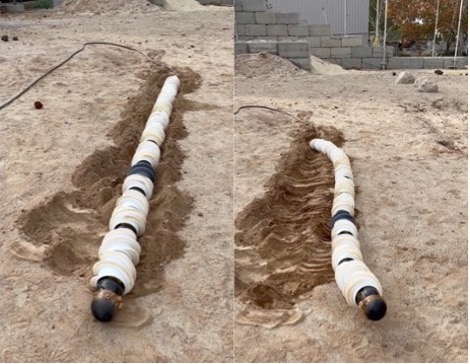}
\caption{EELS robot escaping from a sand trap with a rolling gait.}
\label{fig:eels_left}
\end{subfigure}
\hfill
\begin{subfigure}[t]{0.48\textwidth}
\centering
\includegraphics[width=\linewidth]{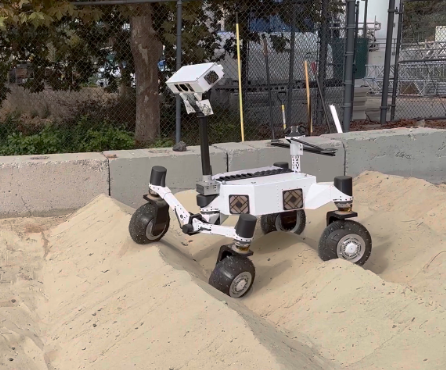}
\caption{JPL’s ERNEST rover with active suspension uses an unconventional walking gait to go over sand dunes}
\label{fig:eels_right}
\end{subfigure}
\caption{Example of high-DOF robots developed at NASA-JPL.}
\label{fig:eels}
\end{figure}

\textbf{High-DOF Robots}: Having more degrees of freedom (DOFs) in actuation than necessary is a great source of adaptivity for planetary robotic systems. A counterexample is Perseverance’s arm. The arm has only has 5 DOFs, enough to place the arm end effector at a 3D position, appropriate for the original mission’s requirements, but does not have enough DOFs to control its roll angle. Later, changes in the Mars Sample Return architecture required Perseverance’s arm to transfer the sample tubes from the rover to the Sample Retrieval Lander (SRL). The 5-DOF design turned out to be a limiting factor for this unplanned use case, which in turn, increased the complexity of SRL’s receiving mechanism ~\cite{Ono2026}. Many mechanical systems flows for space applications have “just enough” complexity for implementing the required functions. For example, conventional rovers have a singular locomotion mode: driving, which works well in nominal conditions. However, the Spirit and Opportunity rovers encountered significant difficulties when trapped in sand. In contrast, robots with redundant DOFs have multiple locomotion modalities. For example, JPL’s snake-like robot, EELS, has 30 DOFs. When it stuck in sand at the Mars Yard, the robot easily escaped by switching to a rolling gait and slithering out sideways (Figure~\ref{fig:eels_left}). This was possible because EELS possesses nearly unlimited locomotion modes, owing to its high numbers of degrees of freedom, and can switch between them via software commands or even improvise new gaits through onboard learning. Similarly, rovers with active suspension joints are shown to be capable of climbing up steeper slopes than passive-suspension rovers by using unconventional gaits such as squirming ~\cite{Bouton2023}, as shown in Figure \ref{fig:eels_right}.

\textbf{Robotic Swarms and Modularity:}
The concept of swarms, modularization, and distributed systems provides a compelling path to the adaptability requirements of PE 3.0. Architectures of modular distributed systems are widely exploited in nature from the microscale (e.g. the cellular assembly of all animals) to the macroscale (e.g. the collaborative social behaviour of organisms such as ants and many species of bees). Because changing the interactions between units in a distributed system can change the form and function of the overall collective, distributed systems can afford a high degree of adaptability that scales beneficially with the number of units and does not require extraordinary complexity nor performance at the level of the individual unit. Practically, these systems also offer favorable properties for robustness—since failure of any single unit does not necessarily imply failure of the collective—and for manufacturability, since repeated production of identical units can reduce cost and concentrate validation effort. The most immediate opportunity for space contexts is in modularizing spacecraft components such that a common set of elementary parts can be repurposed across many missions. This approach is already widely adopted in the design of structural elements e.g. standardized fasteners and aluminum framing. What is lacking are other types of modular elements like unit power sources, computing modules, antennas, and sensors. A longer-term opportunity is the deployment of decentralized spacecraft swarms that can allocate sensing, mobility, and infrastructure roles dynamically as environments and science priorities evolve. 

\begin{figure}[tb]
    \centering
    
    \begin{subfigure}[b]{0.33\textwidth}
        \centering
        \includegraphics[width=\textwidth]{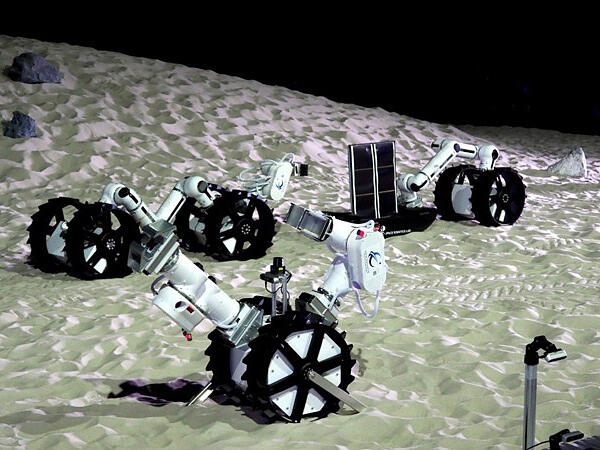}
        \caption{Tohoku University's modular and reconfigurable MoonBot \cite{uno2025moonbot}}
        \label{fig:moonbot}
    \end{subfigure}
    \hfill
    \begin{subfigure}[b]{0.33\textwidth}
        \centering
        \includegraphics[width=\textwidth]{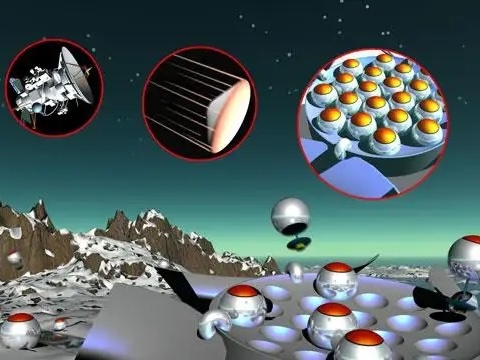}
        \caption{MIT's hopbot swarm \cite{dubowsky2006microbots}\\~}
        \label{fig:mit-hop}
    \end{subfigure}
    \hfill
    \begin{subfigure}[b]{0.33\textwidth}
        \centering
        \includegraphics[width=\textwidth]{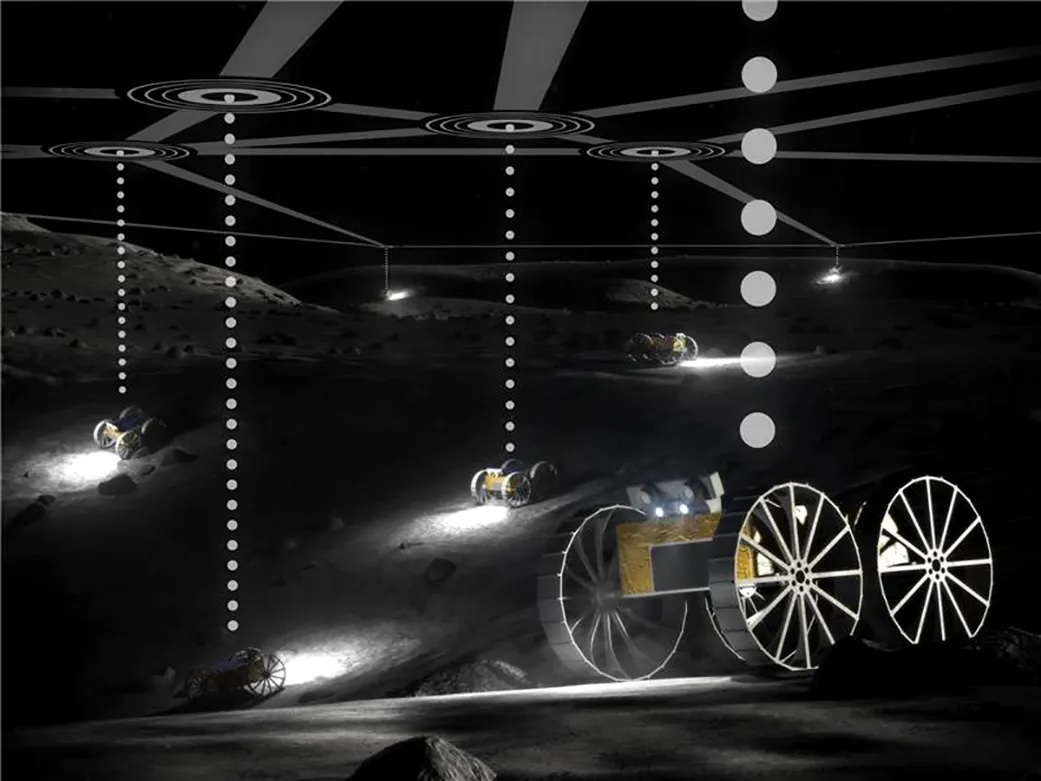}
        \caption{JPL's CADRE mission concept \cite{de2024multi}. Image: NASA/JPL-Caltech}
        \label{fig:cadre}
    \end{subfigure}
    
    \caption{Examples of modular and swarm robot concepts for planetary surface exploration.}
    \label{fig:three_in_row}
\end{figure}

In pursuit of this goal, many groups have developed systems of programmable modular robots that are capable of self-assembly and self-disassembly into different morphologies~\cite{modular1,modular2,modular3,modular4,modular5,modular6,modular7,modular8,modular9}. 
Figure XX shows a few samples from numerous existing studies on modular and swarm robots. 
While the vast majority of them are intended for terrestrial applications, modular and swarm robots are considered for space applications, too. 
For example, MoonBot (\Cref{fig:moonbot}) is a concept of a modular and on-demand reconfigurable robot for performing wide-ranging tasks on the Moon \cite{uno2025moonbot}. 
Previously, planetary subsurface and subsurface exploration by a large-scale swarm of simple hopping robots was proposed (\Cref{fig:mit-hop}) \cite{dubowsky2006microbots}.
JPL's CADRE (Cooperative Autonomous Distributed Robotic Exploration) mini-rover mission, which would deploy three nearly identical rovers on the moon (\Cref{fig:cadre}) \cite{de2024multi}, would be the first humble ``swarm" mission on the planetary surface. 

\begin{figure}[tb]
    \centering
    
    \begin{subfigure}[b]{0.48\textwidth}
        \centering
        \includegraphics[width=\textwidth]{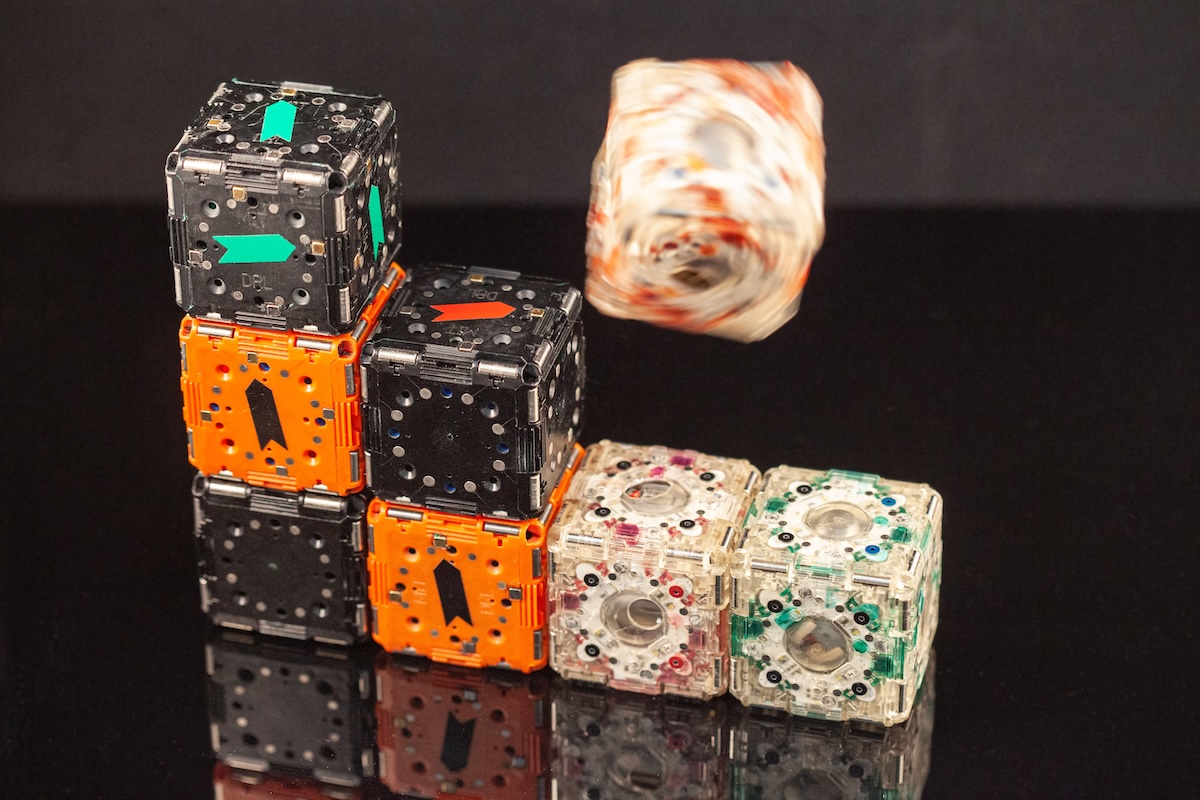}
        \caption{MIT's M-Blocks, terrestrial momentum-driven cubes that form lattice structures \cite{modular3}}
        \label{fig:sub1}
    \end{subfigure}
    \hfill
    \begin{subfigure}[b]{0.48\textwidth}
        \centering
        \includegraphics[width=\textwidth]{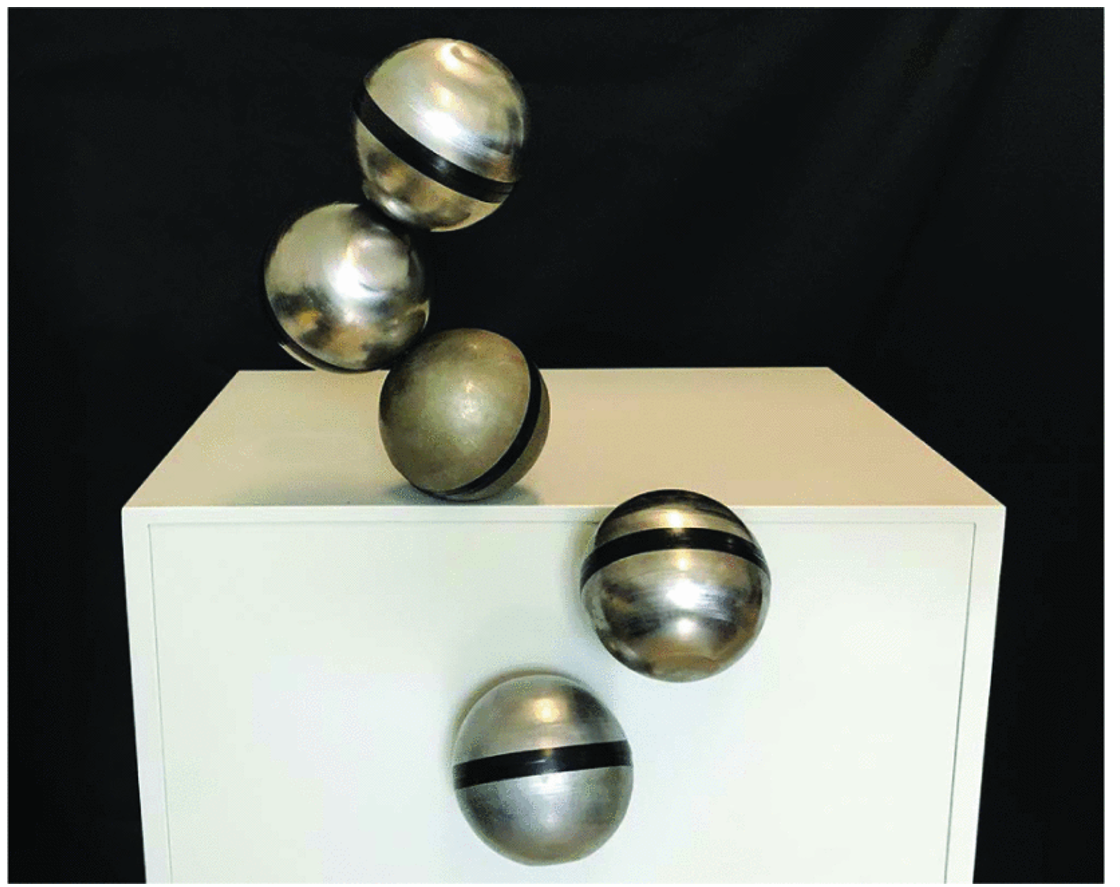}
        \caption{CUHK Shenzhen's FreeBOT, terrestrial spherical robots that connect via continuous magnetic interfaces \cite{freebot}}
        \label{fig:sub2}
    \end{subfigure}
    
    \vspace{0.5em}
    
    \begin{subfigure}[b]{0.48\textwidth}
        \centering
        \includegraphics[width=\textwidth]{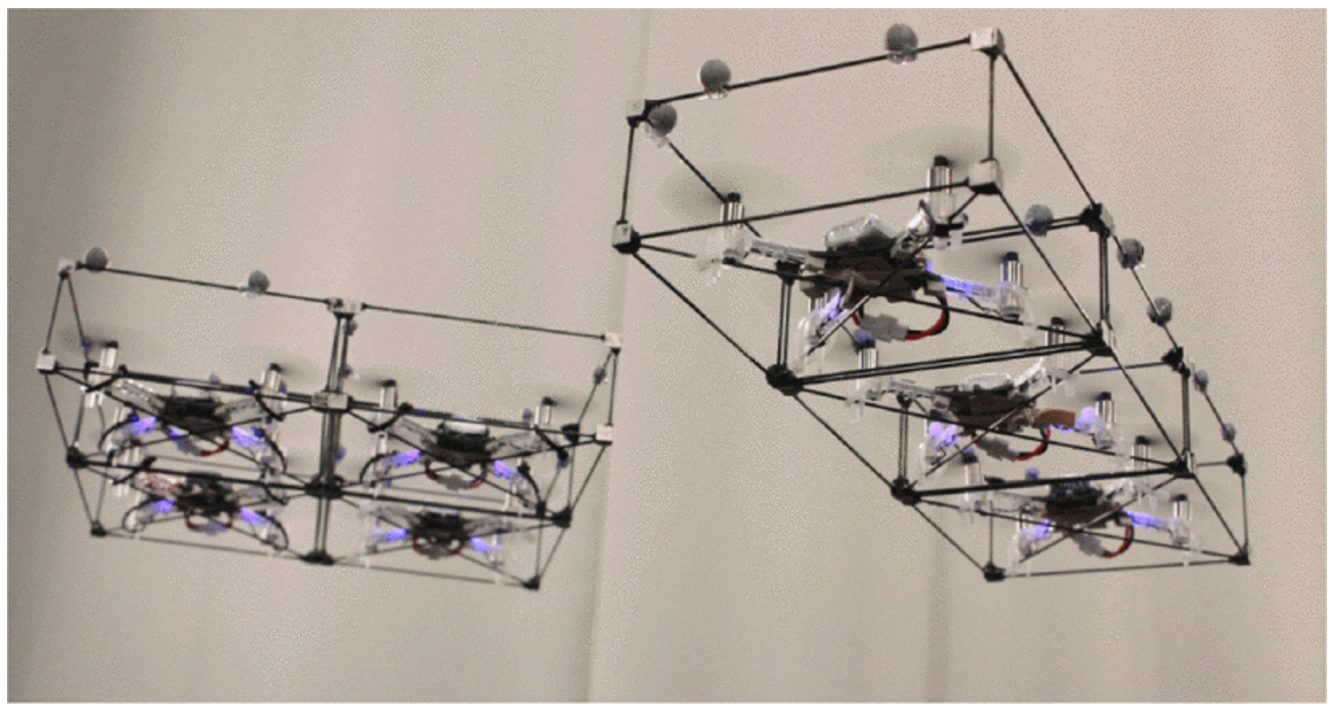}
        \caption{UPenn's ModQuad, aerial quadrotors that connect along rectangular frames \cite{modquad}}
        \label{fig:sub3}
    \end{subfigure}
    \hfill
    \begin{subfigure}[b]{0.48\textwidth}
        \centering
        \includegraphics[width=\textwidth]{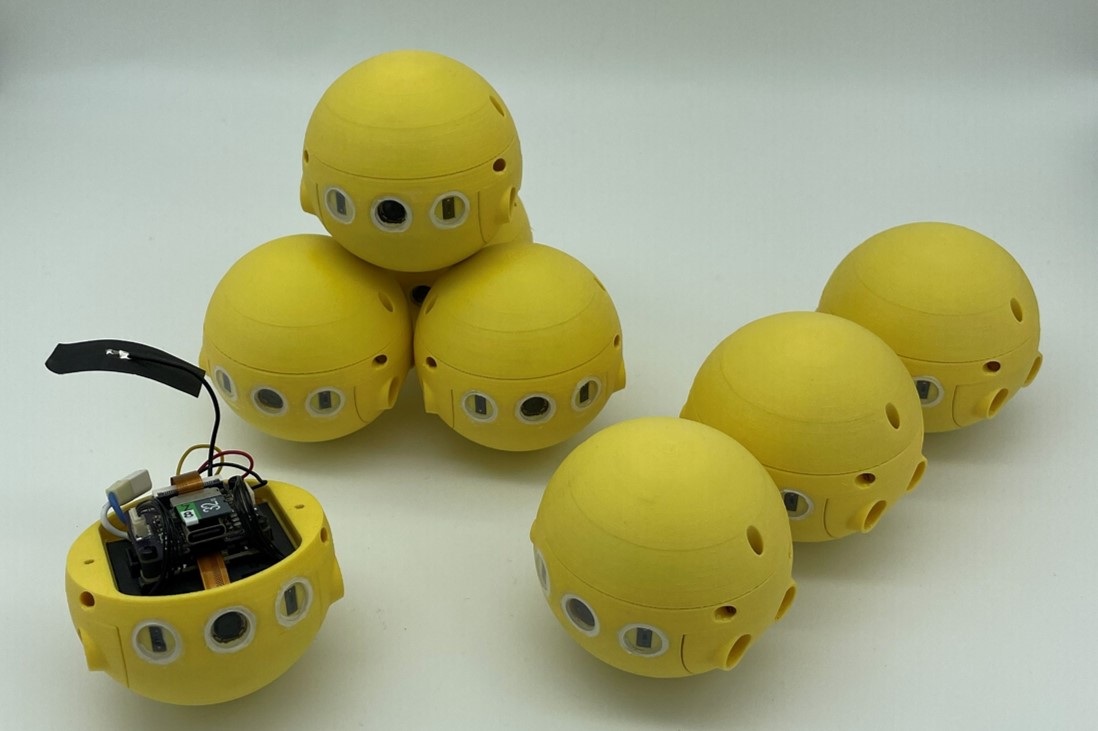}
        \caption{MIT's BµBL, aquatic spherical robots that connect in chain-like structures \cite{modular9}}
        \label{fig:sub4}
    \end{subfigure}
    
    \caption{Examples of modular self-assembling robotic systems.}
    \label{fig:2x2}
\end{figure}

One persistent challenge is in scale; to date, the smallest modular robotic systems exist at relatively coarse millimeter to centimeter scales, which makes hierarchical assemblies of these modules much larger and more massive than comparable, purpose-built machines. There are also significant engineering challenges in equipping each miniature robotic module with sufficient locomotive, sensing, and computing systems to support useful hierarchical structures. Modular hierarchical systems are under-explored in space contexts.

\textbf{Programmable Matter:}
Bringing the concept of modularity to extreme, what if each unit of mass could be made to serve multiple purposes or, better yet, be arbitrarily reprogrammable to serve any purpose? The vision of programmable matter is to bring the concept of computer reprogrammability to physical building blocks of spacecraft, such that the form and function of space-faring machines can be altered \textit{in situ} for radically expanded adaptability and mass reuse. At its most extreme, this concept extends into the far-future ideal of programmable matter, where general-purpose machines might be built from engineered materials with the means to self-assemble and self-disassemble into different forms. 

\subsection{Self-Manufacturing}
\label{sec:self-manufacturing}
In-space manufacturing of spacecraft components is an emerging area relevant to adaptive exploration. In-space manufacturing addresses several constraints; presently, fully assembled spacecraft are expensive to launch from Earth, limited by payload volume, and must be designed to survive ascent loads, leading to structural and system-level features that are unnecessary once in space. Launch availability is also constrained by weather and scheduling, and spacecraft deployed far from Earth are difficult to repair or adapt because any modification requires another launch. These factors bias current spacecraft toward compact, fixed designs that are difficult to scale or reuse. Constructing spacecraft in space or on planetary surfaces avoids these limitations. Systems can be built larger than launch constraints allow, assembled incrementally, and designed for their operating environment rather than peak launch loads.

\begin{figure}[tb]
    \centering
    
    \begin{subfigure}[b]{0.4\textwidth}
        \centering
        \includegraphics[width=\textwidth]{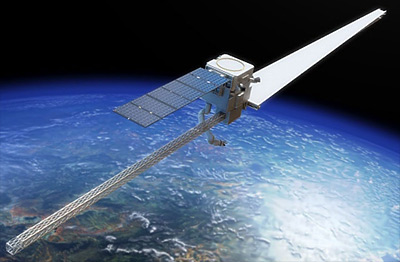}
        \caption{Canceled OSAM-2 mission that would demonstrate in-space manufacturing of 10-meter beams \cite{patane2017archinaut}. Image: NASA/Made in Space.}
        \label{fig:osam2}
    \end{subfigure}
    \hfill
    \begin{subfigure}[b]{0.52\textwidth}
        \centering
        \includegraphics[width=\textwidth]{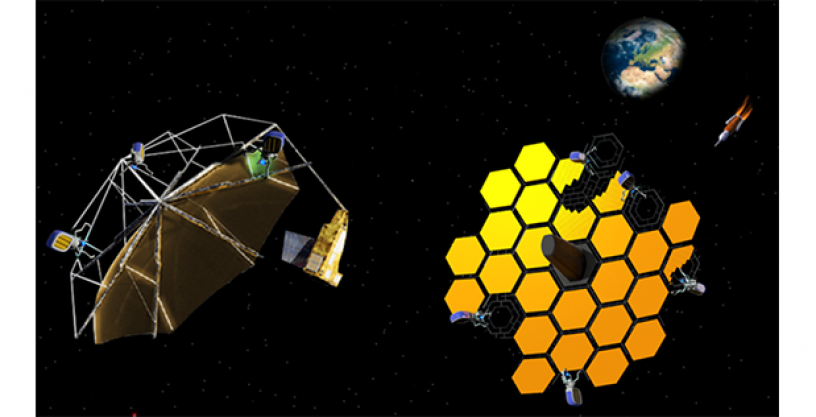}
        \caption{DARPA’s Novel Orbital Moon Manufacturing, Materials, and Mass Efficient Design (NOM4D) concept~\cite{darpaNOM4D2026}. Image: DARPA. \\~ }
        \label{fig:nomad}
    \end{subfigure}
    
    \caption{Examples of in-space self-manufacturing.}
    \label{fig:self-manufacturing}
\end{figure}

\textbf{Additive Manufacturing:} Additive manufacturing (3D printing) has been demonstrated onboard the International Space Station, including fused-deposition modeling of polymers and more recent metal printing efforts using feedstocks launched from Earth ~\cite{ISS_fdm}. These efforts establish that fabrication processes can operate in microgravity and produce functional parts with much faster lead times than manufacturing and launching from Earth. Significant research and development investment has also been put into large-scale free-flying additive manufacturing capabilities that can extrude long truss structures as part of the Archinaut concept for the canceled OSAM-2 (On-Orbit Servicing, Manufacturing and Assembly) mission \cite{patane2017archinaut} (\Cref{fig:osam2}).

\textbf{Modular Assembly:} The concept of modularization explored in the previous subsection is also applicable to on-orbit assembly. There is significant interest in modular spacecraft architectures that enable assembly from generic component feedstocks. Multiple groups have proposed structural and electromechanical building blocks that can be robotically assembled into larger systems ~\cite{armadas, modular_power, freitag2025planning}. These approaches reduce the complexity of in-space fabrication by shifting it toward assembly of pre-manufactured units. There are several planned demonstrations of this technology in the near term. For example, the DARPA NOM4D (Novel Orbital Moon Manufacturing, Materials, and Mass Efficient Design) program plans to test the polymerization of fiber composites and assembly of large apertures on orbit~\cite{darpaNOM4D2026} (\Cref{fig:nomad}). DCUBED plans an on-orbit demonstration of manufacturing of a large truss structure~\cite{DCUBED2023}.

\textbf{In-Space Resource Utilization:} Ultimately, additive manufacturing and modular assembly technologies are limited by their feedstocks: If all materials must be transported from Earth, we still face the fundamental limitation of the rocket equation. To address this, there has also been significant investment over the past few decades in technologies that can harvest materials from other solar-system bodies for manufacturing. Harvesting water and manufacturing propellant has been a particular focus \cite{crawford2015lunar, starr2020mars}. There has also been significant work on harvesting structural materials like aluminum from lunar regolith \cite{zacharias2011real,crawford2015lunar} and semiconductors for manufacturing solar cells \cite{landis2005materials}. Planetary Exploration 3.0 can leverage in-space manufacturing technologies to create structures, instruments, or even entire spacecraft on-demand based on evolving mission demands.
\section{Onboard Intelligence}
\label{sec:intelligence}

Onboard autonomy is a necessary and critical technology for the level of adaptivity required by PE~3.0. 
While there are numerous successful examples of in-space adaptation in the past and current missions, adaptation was typically commanded by ground operations~\cite{Ono2026}. 
Ground-in-the-loop adaptation is limited by communication latency and bandwidth, especially as missions explore outside our Solar System.
In particular, the communication rate is substantially reduced during anomalies, as exemplified by the Galileo and Deep Space 1 anomalies~\cite{Ono2026}.
The only pathway to overcome this limitation is to transfer the decision-making authorities from the ground operations to onboard intelligence.

An analogy is found in biological evolution, where increased levels of intelligence enabled greater levels of adaptability. For example, it is known that the gestures used by many non-human primates for communication are largely shared across species \cite{graham2017gestural,hobaiter2011gestural}, implying that their ``language'' is mostly hard-coded. In contrast, human languages can adapt and evolve because our brain has the ability to assign declarative labels to any objects or abstract concepts. 
Similarly, powerful on-board artificial intelligence should make decisions based on its own stream of experience, respond, and adapt to changes in the environment in real time~\cite{silver2025era}, as well as discover new knowledge.

\subsection{Historical Context}

The evolution of spacecraft autonomy can be broadly understood in three stages.

The earliest spacecraft were fully autonomous in execution but not adaptive. 
For example, Pioneer~4, one of the first U.S. lunar probes launched in 1959, relied on a hydraulic timer to deploy its despin mechanism at a fixed time after launch, and a photoelectric sensor triggered by moonlight. All spacecraft behaviors were hardwired, with no capability for modification once launched. These systems did not even include receivers, precluding any form of command uplink. As a result, spacecraft operation was entirely predetermined, and the role of the ground was limited to passively receiving telemetry. While autonomous in the sense of requiring no human intervention after launch, these systems were fundamentally \textit{non-adaptive}, as they could not respond to unforeseen conditions or be repurposed for unintended objectives.

The introduction of onboard digital computers and command and data handling (C\&DH) subsystems marked the transition to the next stage. 
Spacecraft could now receive commands from the ground, enabling flexible sequencing of activities after launch. This fundamentally changed the paradigm of space operations: spacecraft became less autonomous and increasingly dependent on ground-based interpretation, diagnosis, and decision-making. 
In this operation paradigm, which is characteristic to PE~2.0, adaptation to unforeseen environments and anomalies became possible, but mostly through a human-in-the-loop process. Numerous missions successfully adapted to unexpected situations---for example, through software updates, parameter retuning, or repurposing of onboard assets---but such adaptations relied on ground analysis and commanding \cite{Ono2026}. 
The dependence on the ground operations introduced fundamental limitations, most notably latency and bandwidth constraints in communication, as well as the cognitive bottleneck of ground teams. 

In PE~3.0, we must revert to the level of autonomy of PE~1.0, but in a fundamentally different form. 
Rather than relying on rigid, hardwired behaviors, PE~3.0 envisions \textit{intelligent} onboard autonomy capable of making decisions in response to the observed environment. These systems are not merely executing preplanned sequences, but actively selecting actions, forming and testing hypotheses, and adapting their objectives and strategies \textit{in situ}. Crucially, this autonomy is tightly coupled with software-defined space systems discussed in \Cref{sec:sdss}, which expose hardware-level flexibility that can be exploited in flight. The role of autonomy is therefore not only to control the system, but to fully utilize and expand its capability envelope by reconfiguring and repurposing available resources. 

Recent advances in onboard autonomy provide early steps in this direction. For example, autonomous navigation capabilities on the Perseverance rover have enabled record-setting traverse distances without human review~\cite{doi:10.1126/scirobotics.adi3099}. However, PE~3.0 requires a far more comprehensive form of autonomy that spans science, engineering, and operations. The ultimate vision is that of high-level autonomous agents capable of designing and conducting experiments, prioritizing observations, and managing system resources with limited ground input. Such systems must also reason about their own embodiment, adapting to changes in hardware state by, for example, repurposing sensors, disabling degraded components, or exploiting latent functionality in novel ways. 

Looking further ahead, the full realization of PE~3.0 may include autonomous verification and validation of onboard adaptations, as well as more radical capabilities such as in-space self-reconfiguration and manufacturing. Together, these advances point toward robotic explorers that not only operate independently, but think, learn, and adapt in ways that more closely resemble human scientists and engineers, overcoming the latency and bandwidth limitations inherent in deep space exploration. The following subsections explore these concepts in depth.

\subsection{Onboard Science Data Processing and Planning}

As discussed in \Cref{sec:science-definition}, the science objectives of PE~3.0 will be more dynamic rather than pre-defined, where scientific hypotheses are formulated and measurements are planned on-site. 
Therefore, PE~3.0, will require advanced capabilities to address ambitious scientific goals, creating a strong ``mission pull'' for onboard AI to enable science triage and prioritization of large (multi-Gb) complex data.
Take, for example, a future life detection and life characterization mission, which would employ nanopore sequencers to identify information-containing polymers like DNA and RNA.
The success of such a mission hinges on the ability to rapidly assess and identify the most compelling evidence from high-throughput data streams.  Additionally, mission designs like robotic swarms or melt-probes require that interrelated data must be synthesized and contextualized in real time to uncover high-value targets or environmental anomalies for sampling and analysis.
The mission pull for onboard AI arises from the necessity to meet these demands within the constraints of deep space exploration—limited bandwidth for transmitting data to Earth, restricted onboard power, and the need for autonomous decision-making in environments where latency prohibits real-time intervention from mission control.  
By enabling smart triage and prioritization of scientific data, onboard AI becomes a critical pathway for achieving mission objectives, ensuring the highest-value science is not only captured but acted upon.  This capability empowers missions to adapt dynamically to new discoveries, guiding autonomous operations in dynamic or unknown environments, and ultimately maximizing the scientific return in alignment with the mission's primary goals.

\begin{figure}[tb]
    \centering
    
    \begin{subfigure}[b]{0.49\textwidth}
        \centering
        \includegraphics[width=\textwidth]{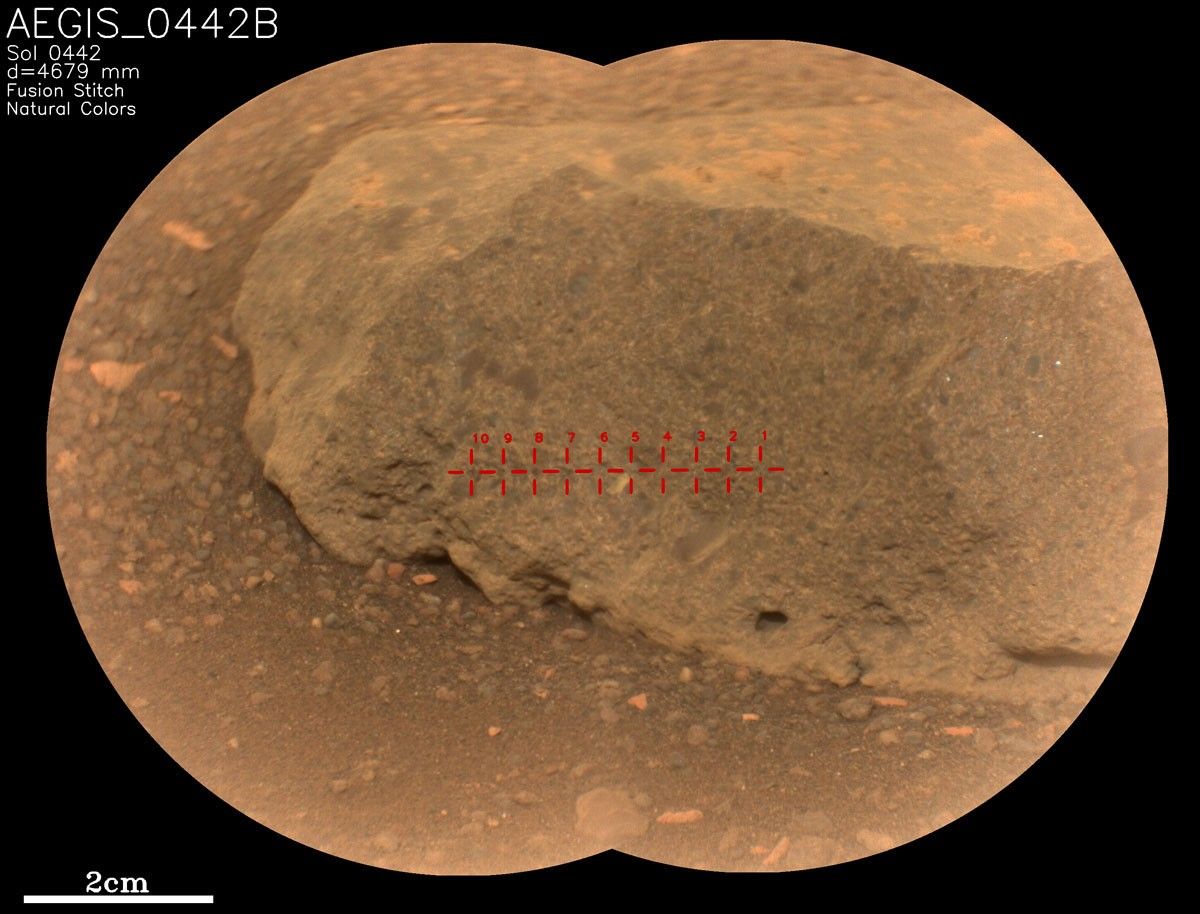}
        \caption{An autonomously selected science target for laser-induced breakdown spectroscopy (LIBS) by AEGIS \cite{Burl2016}. }
        \label{fig:aegis}
    \end{subfigure}
    \hfill
    \begin{subfigure}[b]{0.49\textwidth}
        \centering
        \includegraphics[width=\textwidth]{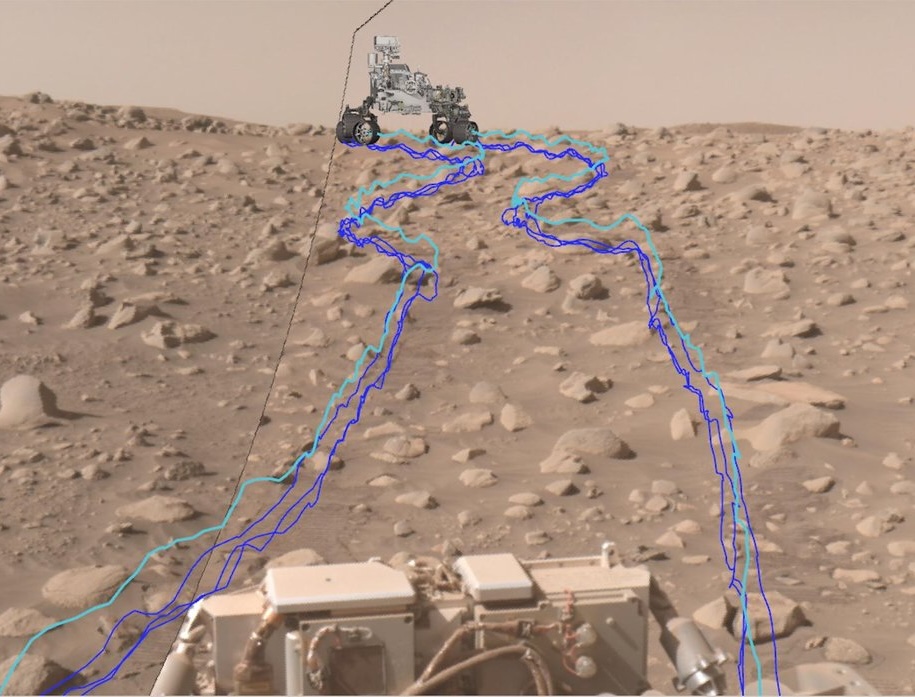}
        \caption{Autonomously planned drive path by Enhanced AutoNav \cite{toupet_perseverance_2026}}
        \label{fig:enav}
    \end{subfigure}
    
    \caption{The Perseverance rover has some of the most advanced onboard autonomy capabilities in space to date. Images: NASA/JPL-Caltech}
    \label{fig:perseverance}
\end{figure}

Autonomous science algorithms have been in development since at least Herb Simon's machine discovery~\cite{Simon1996}.
Laboratory-based autonomous science advanced with the robot scientists Adam and Eve~\cite{Sparkes2010}. Automated science target selection via Rockster and AEGIS (\Cref{fig:aegis}) has been deployed on the Opportunity, Curiosity, and Perseverance rovers~\cite{Burl2016, Verma2023}. 
More recently, experiments such as Ocean Worlds Life Surveyor (OWLS) experiment at NASA Jet Propulsion Laboratory have explored onboard machine learning for adaptive science data selection and prioritization~\cite{Wronkiewicz_2024, 10521206}.
While multiple implementations of science planning with built-in adaptability can be envisioned, further work enabling sophisticated \textit{in situ} science decision making without human operators in the loop is needed. Some adaptive planning algorithms have been tested for this purpose, either by sampling based on expected information or through use of symbolic queries to ask more scientifically-relevant questions \cite{Candela2017,ayton_toward_2020}. Automated task planning and scheduling techniques such as using constraint satisfaction, search, reasoning, and optimization will likely be part of the solution to ensure science return is high while maintaining operational safety and acceptable risk management ~\cite{Candela2017,ayton_toward_2020,Wagner2024DemonstratingPrototype,Russino2025Utility-DrivenMission}. 

Ultimately, it is essential to equip on-board AI systems with the ability to verify their own knowledge, to confidently determine when they are correct, detect when they are wrong, and provide verifiable proof. 
This is especially critical when the AI transmits science data back to Earth: the information must be provably reliable. 
In practice, we should never require an AI agent to assert knowledge it cannot independently verify.

\subsection{Autonomous Navigation}

Future adaptive missions will require spacecraft to sense a target and then interact with it quicker than would be allowed if a ground team were in the loop. 
As a motivating example, imagine a spacecraft that has detected a short-lived cryo-volcanic plume above an icy moon such as Triton.  The desired behavior of an adaptive, autonomous spacecraft would be to respond to the discovery of a plume with a series of actions.  First, the observation would be verified using persistence/validity checks and additional observations, as well as analyzed by onboard intelligence with knowledge of plume characteristics such as predicted spectral, chemical and optical properties.  Then, the spacecraft would use its current knowledge of the encounter trajectory, compute a desired trajectory, and design a maneuver from the current to the desired aimpoint.  Having computed the desired maneuver, and verified that the spacecraft is capable of that $\Delta$V, the maneuver would be checked against appropriate safety and flight rule constraints, and then included in the spacecraft command sequence. Terrain-relative navigation, which used onboard image processing to localize and safely land Perseverance, incorporates some of these aspects of adaptability and autonomy, though the safety of the terrain was defined ahead of time and most of the uncertainty stemmed from localization \cite{johnson_mars_2021}. 
Some smart flyby concepts, such as DARE (Deep-space Autonomous Robotic Explorer) \cite{echigo2025autonomy}, envision sophisticated onboard autonomy and autonomous spacecraft navigation for identifying, targeting, and observing high-value science targets within a quick flyby sequence without ground in the loop. 
Further development along this front could produce algorithms with increasing adaptability and tolerance of higher levels of environmental uncertainty.

Although some form of autonomous optical navigation has been demonstrated on several missions, such as Deep Impact~\cite{Mastrodemos2005}, OSIRIS-REx~\cite{7943684}, and DART~\cite{bellerose2024double}, the specifics of each target/mission application will have to be tested and demonstrated. On the ground, the autonomy software should be tested using actual and simulated images of comets, asteroids, and icy moons. 
In future PE~3.0 missions, in-flight learning will be required since we have no images of bodies such as Triton.  Some of this capability will require integrating adaptability into onboard planning and scheduling algorithms, as the software learns features of the target environment over time through experience.

\subsection{Autonomous Robotic Operations}

Autonomous surface mobility is an essential capability for \textit{in situ} PE~3.0 missions. 
The first extraterrestrial autonomous driving dates back to the Sojourner rover in 1997, although it was limited to reactive obstacle avoidance \cite{mishkin1998experiences}. 
Perseverance, which landed on Mars in 2021, possesses the most advanced autonomous driving capability on Mars to date, and has driven approximately 90\% of its traverse distance autonomously \cite{toupet_perseverance_2026} (\Cref{fig:enav}). 
However, the longest continuous distance driven without ground-in-the-loop remains on the order of hundreds of meters, and the total traverse distance over multiple years is still only on the order of tens of kilometers. 
More importantly, existing autonomous driving capabilities have focused primarily on hazard avoidance and path execution, while more complex maneuvers remain heavily reliant on ground operations. 

For example, recovery from mobility hazards such as sand traps continues to require extensive human intervention. 
During the Mars Exploration Rover mission, Opportunity became embedded in a sand ripple at Purgatory Dune in 2005 and required 39 sols (Martian days) of ground-based analysis, planning, and command sequencing to successfully extract the rover. 
Spirit became permanently embedded in soft soil in 2009 despite months of recovery efforts, ultimately losing its mobility. 
These examples highlight that, while nominal driving can be mostly automated, off-nominal operations remain beyond the capability of current onboard autonomy. 
Terramechanics models could inform automated recovery from sand traps, but they rely on simplifying assumptions and often break down in poorly characterized or extreme terrains \cite{Wong2022}. 
Future systems will require adaptive models that learn from interaction with the terrain through visual and proprioceptive sensing, enabling real-time reasoning about slip, sinkage, and traction. Early work on learned or adaptive terrain models suggests a path forward, but robust, flight-ready implementations remain an open challenge~\cite{Lupu2025}.

In contrast to mobility, the level of onboard autonomy for other robotic operations—such as manipulation, instrument placement, and sampling—is significantly lower. 
These activities involve complex contact interactions, tight clearances, and high-consequence failure modes, and are therefore still dominated by ground-in-the-loop planning and verification. 
For example, in the Perseverance mission, collision checking of the robotic arm is performed on the ground using detailed models and simulation tools prior to execution. 
As missions extend to more extreme environments, such as subsurface ocean worlds, the challenge becomes even more acute. Autonomous subsurface mobility—potentially in confined, unstructured, and poorly sensed environments—will require capabilities far beyond current practice.

A fundamental limitation of current planetary robotic autonomy is that it is largely model-based and preconfigured. 
While parameters can be tuned and flight software can be updated from the ground, the onboard system itself does not adapt in real time to fundamentally new conditions. 
This paradigm is insufficient for PE~3.0 missions, where spacecraft must operate in previously unseen environments with limited or no opportunity for ground intervention. 
Future systems will need to perform on-site learning, rapidly building and updating models of their environment and their own interaction with it. 
Such capabilities would enable robots to adapt their behaviors—e.g., mobility strategies, manipulation approaches, or sampling techniques—in response to unexpected conditions. 
However, achieving this level of adaptivity in a reliable and safe manner remains a major challenge, requiring advances in learning-enabled control, uncertainty quantification, and onboard verification and validation.

\subsection{Controls and Embodied AI for Software-Defined Space Systems}

A central requirement for PE~3.0 is the ability to fully exploit the hardware-level flexibility of software-defined space systems through onboard control. 
While reconfigurable hardware and multi-functional components provide the \textit{potential} for adaptability, it is the control system that realizes this potential in operation. 
In particular, control must not only execute pre-defined behaviors, but dynamically adapt system behavior in response to changing or unforeseen conditions, including environmental variability, component degradation, and failures.

To achieve this, control systems must evolve beyond fixed, model-based formulations toward adaptive and learning-enabled approaches. 
Traditional control and planning algorithms rely on models of system dynamics and environment interactions that are defined pre-flight and remain largely static. 
However, in long-duration missions and poorly characterized environments, these models inevitably become inaccurate due to factors such as wear, damage, and unmodeled terrain properties~\cite{doi:10.1126/scirobotics.adn4722}. 
As a result, PE~3.0 systems require mechanisms to continuously update their internal models based on onboard observations, ensuring that control and planning remain valid over time~\cite{doi:10.1126/scirobotics.abm6597}. 
While such model updates can in principle be performed on the ground, the latency and bandwidth constraints of deep space missions make onboard adaptation the preferred and often necessary approach.

Beyond adapting control policies, PE~3.0 systems must also reason over a significantly expanded space of possible system configurations enabled by software-defined architectures. 
This includes selecting among multiple operating modes, reconfiguring subsystems, and repurposing hardware to achieve mission objectives under new constraints. 
Control, in this sense, becomes a higher-level decision-making process that integrates perception, system identification, and action selection, tightly coupling autonomy with the underlying system flexibility.

Embodied AI provides a framework for addressing these challenges by jointly reasoning about control, perception, and physical embodiment. 
Highly adaptive systems often possess substantial physical degrees of freedom (DOFs), enabling multiple locomotion and manipulation strategies. 
For example, robots that combine locomotion modalities such as walking and flying~\cite{doi:10.1126/scirobotics.abf8136} or driving and flying~\cite{sihite2023m4} illustrate how additional DOFs can expand the achievable behavior space. 
However, effectively utilizing such flexibility is a major challenge. 
Biological systems provide a useful reference: humans have on the order of 200 DOFs, and some snake species exceed 800 DOFs~\cite{snake_dof}, yet they achieve robust and adaptive behavior across diverse environments. 
In contrast, most robotic systems operate with far fewer DOFs and rely on relatively rigid control strategies.

Existing control approaches such as model predictive control and reinforcement learning have demonstrated promising results in domains including autonomous driving, legged locomotion, and dynamic manipulation~\cite{Urmson2008, Miki2022, Bhardwaj2022}. 
However, these methods are typically applied within constrained operating regimes and depend heavily on pre-collected data or carefully tuned models. 
Extending them to the open-ended, uncertain environments of PE~3.0 requires advances in online learning, generalization, and robustness.

Another key limitation of current embodied AI systems is their reliance on a narrow set of sensing modalities, primarily vision and proprioception. 
Biological systems, in contrast, leverage rich multi-modal sensing, particularly tactile feedback, to interact with complex environments. 
For example, the human hand contains tens of thousands of mechanoreceptors, and animals such as the star-nosed mole rely heavily on tactile sensing for rapid and precise interaction with their surroundings~\cite{CATANIA2005R863}. 
While artificial tactile sensing has been studied extensively in the context of grasping and manipulation~\cite{Cutkosky2008}, its application to locomotion and terrain interaction remains limited~\cite{8823987}. 
Similarly, while AI has achieved remarkable progress in visual and auditory perception, the interpretation and integration of tactile data remain relatively underdeveloped~\cite{doi:10.1126/scirobotics.abc8134}. 

Advancing embodied AI for PE~3.0 will therefore require not only improved control and learning algorithms, but also richer sensing and tighter integration between perception and action. 
In particular, the development of systems that can learn from physical interaction—combining visual, proprioceptive, and tactile feedback—offers a promising path toward robust, adaptive control in unknown environments. 
\section{PE 3.0 Mission Concepts}
\label{sec:missions}

We explore concept studies for three notional adaptive one-shot missions enabled by software-defined space systems: (1) a Neptune/Triton smart flyby, (2) an icy-moon orbiter-lander with surface-to-subsurface access capabilities, and (3) an Oort Cloud mission. By considering the potential implementation of PE 3.0 in these compelling scenarios, we highlight key advantages and identify areas of technology development needed to make such concepts a reality.

\subsection{Neptune/Triton Smart Flyby}
\label{sec:neptune}

A mission to Neptune and its satellite system is of great interest to the science community~\cite{NASEM2023}. Adaptivity is desirable because our knowledge of the system is quite primitive at this time.  Early observations will provide information such as brightness and spectral features that can be improved by onboard tuning of remote sensing instrument settings.  Additionally, flybys of Triton, the largest Moon of Neptune and a likely captured Kuiper Belt object, will benefit from on-board adaptability, particularly from autonomous navigation-based targeting of cryovolcanic plumes~\cite{Hofgartner_2022}. By flying through the plumes, \textit{in situ} sampling can augment remote sensing data to determine species that are present in the erupted material.  Probes carried by the spacecraft will also benefit from adaptability during the mission, allowing entry/impact targeting to benefit from precise navigation prior to release, and from improved observation parameters informed by the latest remote sensing data.

\paragraph{Scientific Objectives}
The mission goals for both Neptune and Triton are shown in~\Cref{tab:neptune}. The goals listed for Neptune would be achieved through observations during the entire mission, with remote sensing data complemented by precisely targeted atmospheric probes.  For Triton, the mission would include multiple flybys, similar to the Cassini tour or the plannet Europa Clipper flybys.  The mission would also include multiple Triton probes, as well as remote sensing instruments and \textit{in situ} sampling instruments to be used during atmospheric and plume fly-throughs. 

\begin{table}[htb]
\caption{Mission goals and science objectives for the Neptune/Triton flyby concept.}
\label{tab:neptune}
\centering\small
\begin{tabular}{@{}p{2.0cm}p{5cm}p{5.5cm}@{}}
\toprule
\textbf{Body} & \textbf{Science Goal} & \textbf{Candidate Payload} \\
\midrule
Neptune & Internal structure & Gravity field, magnetic field \\
Neptune & Atmosphere structure and dynamics & Atmospheric probe with mass spectrometer (MS), microwave radiometry \\
Neptune & Origin in the protosolar disk & Atmospheric probe with MS (isotopes, noble gas) \\
Triton & Is Triton an ocean world? & Gravity field, magnetic field \\
Triton & Plume composition and processes & Fly through with MS, remote sensing instruments, \textit{in situ} \\
Triton & Ice shell subsurface structure and processes & Ground penetrating radar, seismometry, surface/subsurface explorer \\
\bottomrule
\end{tabular}
\end{table}

\paragraph{Concept of Operations and Architecture}

After a multi-year cruise to Neptune, the spacecraft would be inserted into a large retrograde capture orbit, and subsequently adjusted to raise periapsis to a safe altitude and place apoapsis at several times the Triton distance.  A retrograde orbit is chosen to minimize the flyby velocities at each Triton flyby, since it is also retrograde.  The placement of apoapsis is chosen at a non-integer multiple of Triton's 5.8 day period, so that the flybys occur at different Neptune longitudes.  Since Triton’s orbit is synchronous with its rotation, this also allows for a variation in the Triton-centered longitude of the various flybys.  

As shown in~\Cref{tab:neptune}, the spacecraft would carry probes for close measurements of both bodies.  The number and sequence of probes will be decided in a future architectural study.  Probes would be used for characterization of the atmospheres.  For Neptune the probe transmission would continue as deep into the atmosphere as possible to provide clues to the internal structure of the planet.  For Triton, the probes would be used to measure atmospheric structure during descent, but also would be designed to survive landing on the surface.  By using multiple Triton probes, the various terrain types, including polar caps, the so-called ‘Canteloupe Terrain,’ and presumed cryovolcanic plains would all be observed separately.  

Both bodies would be studied from orbit, with observation time being shared to optimize the information for each.  Repeated autonomous flybys of Triton would be used for more detailed remote sensing, and after initial reconnaissance is completed, for low altitude flybys through any volcanic plumes that may be observed.  The plume flybys would use \textit{in situ} measurements of the plume material to characterize the outflow, currently assumed to be a combination of gases, water, and dust.  

\paragraph{Adaptivity Needs and Benefits}
Adaptability would be used by this mission in several ways, through both instrument and spacecraft system-level autonomy.  Instrument autonomy would be used to learn from early measurements and adjust follow-on observations towards more specific compounds.  For example, detection of complex hydrocarbons on a flyby would be tuned in a future flyby to specific amino acids or fatty acids which might indicate biological activity.  Because plumes could potentially last for many months, this capability could be used to move systematically through a range of molecules of interest during a sequence of flybys.  Here we describe how the spacecraft would execute a Triton plume encounter, but many of the same techniques would be used for a probe targeting/release event or a Neptune close approach.

The spacecraft systems would cooperate in the targeting of each flyby and would take advantage of onboard capability.  Combining prior ephemeris knowledge with observations of the target, the software would determine a modification to the trajectory and then use spacecraft propulsion to effect that change.  Advanced autonomous orbit determination using optical images would compute optimum flyby conditions.  Onboard capability is required because the resolution required for targeting a flyby is achieved with less time-to-go than would be required by a ground solution.  The system would then compute and execute the maneuvers required to achieve the desired flyby condition (altitude, flyby time to intersect the plume).  

Following the completion of an updated target for the flyby, the on-board observation sequence and instrument observation parameters would be calculated and executed during the flyby.  The spacecraft would also autonomously tag and store all of the encounter scientific data, to be used both for downlink and future observation planning.  

As mentioned, a similar approach would be used for other observation sequences.  In the case of a probe entry, the spacecraft would autonomously compute and execute attitude maneuvers to target and release the probe, then autonomously compute and execute a deflection maneuver to ensure that the spacecraft does not follow the probe into the target body.  In addition to the data recording required for a flyby, a probe sequence would also require the telecommunications systems between the spacecraft and probe to be configured and pointed properly for successful crosslink of the data.  

Although a Neptune/Triton mission could theoretically be performed without onboard adaptability, the resulting design space for each observation would be much wider.  This would result in fewer guaranteed plume fly-throughs, less precise targeting for each probe entry, and markedly less data from the overall mission.  For a distant target such as the Neptune system, where a follow-on mission would be at least tens of years into the future, on board adaptability ensures rich scientific return and the answer to most of the currently known questions about the Neptune system in only one-shot.

\subsection{Ocean World Explorer}

On Earth, diverse microbial habitats exist, and the microbial community size and diversity depend heavily on the habitability of the environment, which is nonuniform ~\cite{Cockell2025, Ratliff2023, Bell2008}. This leads to the development of distinct microbial biogeographies in Earth environments which differ from small scales (\textasciitilde{}1m) in sediments to global scales. Unlike static lander concepts such as the Enceladus Orbilander ~\cite{MacKenzie2021}, which may follow a predefined operational timeline, an astrobiology mission with adaptive systems would thrive on iteration and feedback, allowing a more nuanced study of habitability and life processes across numerous environments. Large uncertainties persist about icy ocean world environments as subsurface liquid regions and their access points are not amenable to study by remote sensing. These uncertainties will continue to persist until the first exploration systems arrive there. Understanding the spatial distribution of biosignatures at multiple scales is critical to the success of astrobiology-focused missions, as demonstrated by analog studies showing significant variation even in apparently homogeneous environments~\cite{doi:10.1021/acsearthspacechem.1c00390}.

\paragraph{Scientific Objectives}
By autonomously analyzing plume, surface, and subsurface materials for biosignatures in real time, adaptive systems could detect promising indicators of life and adjust mission goals to probe deeper for more complex biological patterns across an icy ocean world~\cite{Howell2025}. For example, the life detection suite would be able to create a comprehensive set of measurements of life within its environments, not just limited biosignature detection capability based on a singular sample type and molecule class~\cite{MacKenzie2022}.

As we robotically explore diverse habitats and characterize life that may exist within these diverse environments, we will construct the first microbial geographies of a world beyond Earth. Ultimately, this ability to integrate findings across spatial and temporal scales could push the mission's scope to include not only whether life exists, but what kind of life it might be, how it thrives in an icy ocean world’s unique conditions, and its implications for life elsewhere in the universe. This will fundamentally transform our understanding of life and its limits.

\paragraph{Concept of Operations and Architecture}

\begin{figure}[tb]
\centering
\begin{subfigure}[t]{0.57\textwidth}
\centering
\includegraphics[width=\linewidth]{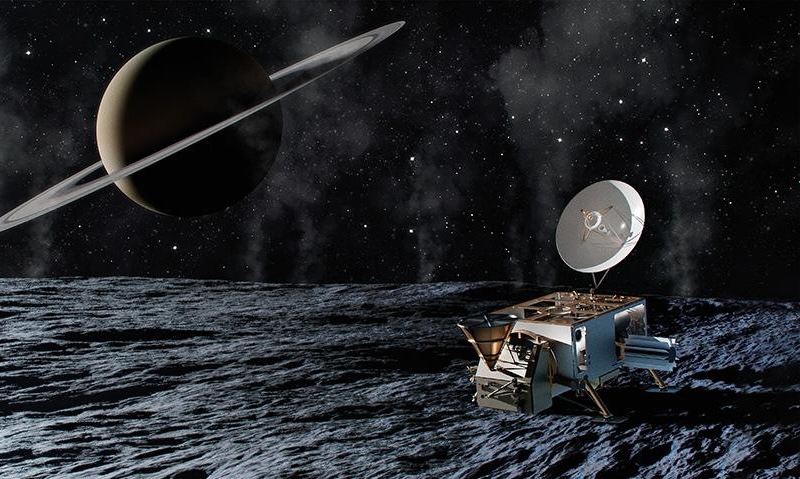}
\caption{Enceladus Orbilander concept~\cite{MacKenzie2022}. NASA/JHU-APL.}
\label{fig:orbilander}
\end{subfigure}
\hfill
\begin{subfigure}[t]{0.41\textwidth}
\centering
\includegraphics[width=\linewidth]{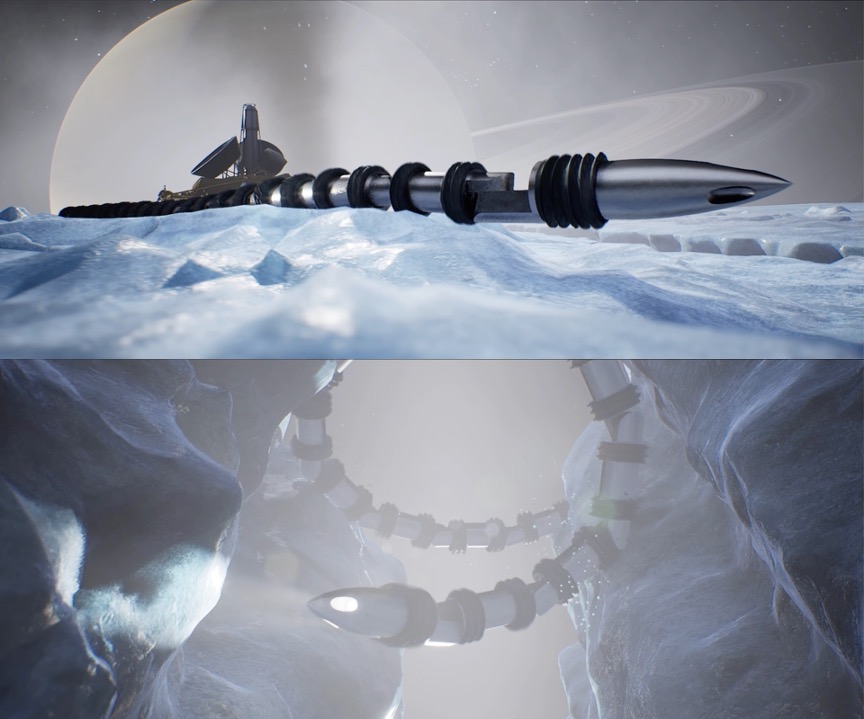}
\caption{EELS concept~\cite{vaquero2024eels}. NASA/JPL-Caltech.}
\label{fig:eels_concept}
\end{subfigure}
\caption{Existing mission concepts that inspired the Ocean World Explorer concept.}
\end{figure}

This adaptable mission is a combination of ideas inspired by the Europa Lander~\cite{hand2022science}, Enceladus Vent Explorer~\cite{ono2018enceladus}, Enceladus Orbilander~\cite{MacKenzie2021} (\Cref{fig:orbilander}), and the EELS concepts~\cite{vaquero2024eels} (\Cref{fig:eels_concept}). An Enceladus exploration version of this mission would launch from Earth, travel to the Saturnian system (7 year cruise) and enter its orbit, spend 4.5 years operating in Saturn orbit, enter orbit around Enceladus, conduct 1.5 years of science and mapping in Enceladus orbit to identify future surface study locations, land on the Enceladus ice sheet, spend 1.5 years exploring Enceladus’ surface, and finally descend into the sub-surface ocean for 6 months of subsurface ops. Launch to landing is about 13 years if we use the Orbilander timeline as a model~\cite{MacKenzie2021}. 

These mission phases are characterized by markedly different spacecraft and scientific demands, necessitating a flexible and adaptable spacecraft with versatile hardware. One key aspect to ensure the spacecraft architecture does not grow prohibitively large is to minimize mission-phase-specific hardware. For example, a single communications system that could be used across all aspects of the mission or a camera system that could serve as a star tracker, mapping camera, and surface imager would save mass and complexity. Additionally, a sample processing instrument that could accept samples from the plume, ice, and the subsurface ocean would be a necessity, albeit with unique sample acquisition methodologies.

\paragraph{Adaptivity Needs and Benefits}
This mission architecture must address the multiple environments in which life-seeking and characterizing measurements may be made (low-altitude orbit of Enceladus for sample collection in the upper plume, collections on the icy surface both near and far from vent openings, in-ice collection of samples, both near and far from the plume vents, and subsurface ocean samples both near the ice crust and near the ocean floor) and the multiple modes of sample collection and mobility which would be needed. The mobility needs on Enceladus, or any ocean world where penetration through an ice shell is desired, are diverse~\cite{Chodas2023}. Challenges are raised from each specific area, from orbital surveying of plume(s), to surface operations and sampling, to descending through vents or melting through ice to reach the ocean below, to adapting to high pressures and approaching the dark low-energy seafloor. One challenge will be conserving some design elements between all of these locations in order to reduce mission mass, complexity and cost. A minimum amount of mission-phase-specific and environmental-specific hardware is desired, making modular reconfigurable systems ideal, such as bots which have rotating screw-like fins along the exterior and can connect with other bots in the swarm. Screw-like fins may act as a high-traction wheel or snake when rolling along the surface, a screw when burrowing or melting through tight openings of the ice, and a propeller when swimming through the ocean. Another innovation is the ability for bots to physically connect, transfer power and information, and act as one larger bot using a control surface designed with all three environments in mind.

Getting information from the deep subsurface to the surface and then to Earth will require an information relay station on the ice shell surface to transmit information to an overhead orbiter or directly to Earth. Similarly, tethered power from a surface station, generated by nuclear fission reactors, may be the optimal choice to provide the large amounts of adaptability and optionality needed to explore various areas quickly both undersea and throughout the ice shell~\cite{Gibson2018}. A compact fission reactor could sustain power for many decades by controlling how fast its fuel is used. Early in the mission it could operate at low output to conserve fuel, then ramp up for demanding phases like melting through ice or powering surface systems. This would make the mission far longer-lived and more flexible than one supported by one or more Radioisotope Thermoelectric Generators (RTGs). This could occur on a surface station and be tethered down to a less complex probe below the ice. This station may deploy mobile systems or build them opportunistically with additive or subtractive manufacturing in order to adapt to the environment.

Sample capture on Enceladus demands adaptability given uncertain vent conditions. We imagine several new complementary strategies: a trombone-architecture tent placed over active vents to decelerate and collect plume material; absorbent ``water capture balls'' deployed during quiescent periods to sink toward the subsurface ocean and be expelled upon jet resumption. These, paired with supplementary approaches including sheet collectors, excavation systems, and robotic probes (e.g., EELS, melt probes) could provide 2–3 redundant methods to ensure robust sample collection across unexpected surface and subsurface conditions.

Certain instruments could perform analyses in a nondestructive fashion (e.g., infrared or Raman spectroscopy, contactless conductivity detection), while others consume/destroy the sample in order to perform the measurement (e.g., mass spectrometry). To enhance science return, it would be beneficial to couple non-destructive and destructive techniques into a single in-line instrument suite~\cite{Mora2022}. Examples of this already exist, such as integration of capillary electrophoresis (CE) with laser-induced fluorescence (LIF), capacitively coupled contactless conductivity detection (C4D) and electrospray ionization mass spectrometry (ESI-MS) into a single instrument~\cite{Mora2022,brinckerhoff2022europan}. Methods like capillary electrophoresis are robust to challenging sample matrices, allowing numerous liquid sample types to be analyzed by the same method \cite{seaton2023robust}. This may not necessarily be done prior to launch, but if interfaces could be standardized between different instrument systems, such combinations could be made on arrival informed by preliminary measurements of sample composition (such as conductivity as a proxy for salinity, or initial plume flythrough measurements constraining volatile and organic compositions). Standardized interfaces could also be utilized for swapping out sampling systems (plume sampler versus surface ice scoop versus sipper of ocean liquids) to deliver samples to a given instrument suite at different mission phases (orbit, surface operations, subsurface operations).

\subsection{ORIGIN: Oort Reconnaissance for Interstellar Genesis and Inheritance}

The need for adaptable spacecraft and one-shot mission concepts that can conduct exploratory science and follow up with hypothesis testing is most acute in the outermost reaches of the solar system and beyond. 
The vast distances prohibit iterative missions building on what each observed before, and the lag in communications, particularly as a spacecraft approaches another solar system, necessitates decision-making without humans-in-the-loop.  
We therefore consider an Oort cloud mission as a representative case, as this region may hold key information about solar system formation, evolution, and interaction with extrasolar material, and may, perhaps, serve as a final resupply point for future interstellar missions.

\begin{figure}
    \centering
    \includegraphics[width=0.6\linewidth]{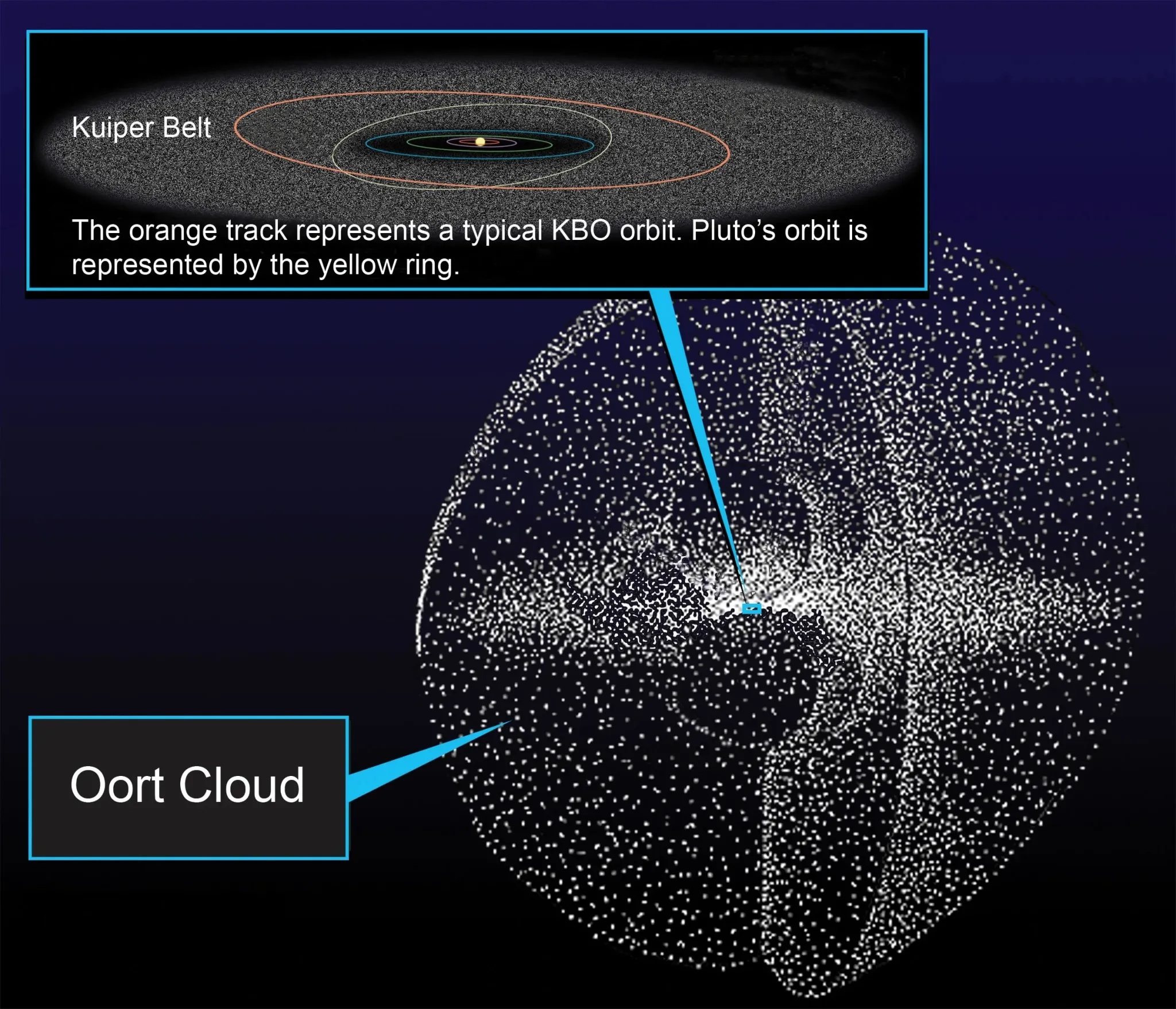}
    \caption{Hypothesized structure of the Oort Cloud. Image: NASA}
    \label{fig:oort}
\end{figure}

\paragraph{Scientific Objectives}
The Oort cloud is the theorized source of long-period comets.
It is thought to contain billions of planetesimals in a roughly spherical distribution, with a subset (the Hills cloud) closer to the ecliptic, as shown in \Cref{fig:oort}. 
The inner edge of the Oort Cloud is thought to be located between 2,000 and 5,000 AU from the Sun, with the outer edge being located somewhere between 10,000 and 100,000 AU, or 0.16 and 1.6 light-years, from the Sun.
The outer Oort cloud is only loosely bound to the Sun, marking the cosmographic boundary of the solar system. 
Although these bodies are hypothesized to have formed closer to the Sun and been ejected outward, they remain unobservable with current instruments. 
Thus, exploration of the Oort cloud is a prime example where adaptive missions must enable both exploratory and hypothesis-driven science. 
Even confirming its existence would be high value; further constraining its composition, age, and origin would yield transformative insights into the origins and evolution of the solar system. Much remains unknown due to low albedo, extreme distance, and low object density, leading to large separations between bodies. 
While comets provide occasional samples of Oort cloud material, it is unclear whether they are representative. 
It is also unknown whether Oort cloud bodies primarily originate from the protosolar nebula or include substantial extrasolar material. 

Given these uncertainties, an initial objective is to catalog object distribution, size, and dynamics through imaging, albedo measurements, and tracking. 
Spectroscopic observations across a wide wavelength range (UV to far IR) would enable compositional analysis, including water, silicates, and organics. 
Volatile abundances and isotopic ratios are critical for determining origin, while dust analyzers (e.g., SUDA-like instruments \cite{Kempf2025}) and solar occultation techniques could characterize fine-grained material.

Due to the limited prior knowledge, we frame the science around two primary questions with associated subquestions:
\begin{enumerate}[label=\arabic*.]
    \item What is the origin of the Oort cloud?  Its distance, slow orbits, and lack of constraint to the elliptic plain means it could have formed with the solar system from the same protosolar nebula, or be extrasolar material from beyond. It could be a mix, but the relative ratios and how that material is distributed are also of interest. To address this question, we must address subquestions:
    \begin{enumerate}[label=\alph*.]
        \item What are the abundances and isotopic compositions of volatiles? Comparison with primitive solar system bodies will constrain whether the material is solar or extrasolar in origin.
        \item What are the sizes and trajectories of Oort cloud bodies? Local dynamical structure can provide insight into formation and evolution processes.
    \end{enumerate}
    \item Can the Oort cloud support \textit{in situ} resource utilization? The Oort cloud represents a last stop on the way out of our solar system to interstellar destinations where existing bodies could be used, whether for gravity assists (if there is a Planet X / Nine) or for collecting material to be used during the continuing phase of an interstellar mission. 
    \begin{enumerate}[label=\alph*.]
        \item What molecular volatiles are abundant? Species such as hydrogen, water, and organics are relevant for propulsion, life support, and in-space manufacturing, while also informing formation history.
    \end{enumerate}
\end{enumerate}

\paragraph{Concept of Operations and Architecture}

The mission begins with with a long cruise to the outer solar system and orbit circularization within the Oort cloud. 
A key challenge is selecting the optimal orbital distance given the highly uncertain distribution, size, and composition of Oort cloud objects, requiring significant adaptivity. 
A propulsion system capable of large orbital maneuvers is needed, with nuclear propulsion as a candidate. 
The mission operates in two primary modes: broad exploration (e.g., remote sensing and spectroscopy) and targeted characterization via autonomous target selection. 
These modes may occur sequentially, concurrently, or iteratively as knowledge evolves.

The architecture is centered on a mother–daughter spacecraft system. 
A primary mothership conducts initial reconnaissance, including flybys to refine target selection, and serves as the command, communication, and sensing hub. 
It deploys multiple heterogeneous daughtercraft for targeted investigation of selected bodies. 
These daughtercraft may include orbiters, landers, impactors, or dust probes, enabling adaptation to diverse environments. 
The mothership focuses on exploratory science under high uncertainty, while daughtercraft perform more specialized, hypothesis-driven measurements as environmental knowledge improves.
The system can be extended to include multiple motherships, increasing resilience, spatial coverage, and parallel science operations. 
Even if a mothership fails, others can continue the mission. 
The architecture emphasizes reconfigurability and modularity: the mothership hosts high-power, multi-functional systems such as a laser enabling orbital LIBS/Raman spectroscopy, power beaming, and optical communication with Earth, while daughtercraft carry modular, standardized instruments (e.g., spectrometers, mass analyzers, surface science payloads) connected through common interfaces to enable reconfiguration and repair.

Instrument selection must balance adaptability with resource constraints (mass, power), leveraging affordance-based design. 
This includes maximizing the range of science achievable with individual instruments and enabling flexible combinations of multifunctional payloads as science objectives evolve during the mission. 
Together, the mother–daughter system enables adaptive exploration across both unknown and increasingly characterized environments.

\paragraph{Adaptivity Needs and Benefits}
Adaptive systems operating in highly unknown environments, such as the Oort Cloud, must balance exploration and exploitation. 
Exploration seeks to gather information about unknown regions of a state or parameter space, while exploitation focuses on actions expected to yield high-value outcomes based on current knowledge. 
This tradeoff is well studied in machine learning, for example in the multi-armed bandit problem, where an agent must choose among actions with uncertain and time-varying rewards \cite{Audibert2009}. 
Similar challenges arise in robotics, such as motion and path planning in partially known environments, where robots must decide whether to explore new regions or focus on areas already known to be informative \cite{Rickert2008, Marchant2014}. 
In all cases, adaptability enables more effective learning of the underlying information or value distribution over time. 
Pure exploration is inefficient, while pure exploitation risks missing higher-value opportunities.

As discussed in \Cref{sec:intro}, PE~2.0 addresses this tradeoff across multiple missions: early missions emphasize exploration, followed by hypothesis-driven missions that exploit accumulated knowledge. 
In contrast, the ORIGIN mission must perform both exploration and exploitation within a single mission, as the relevant science questions cannot be fully defined \textit{a priori}. 
Moreover, the mission will likely transition repeatedly between these modes as discoveries are made and hypotheses are tested at specific Oort cloud bodies. 
This necessitates adaptable alternatives to the traditional science traceability matrix (STM), as discussed in \Cref{sec:science-definition}.

Enabling this capability requires onboard autonomy for scientific decision-making. 
The system must determine which measurements to perform, on which targets, and at what time. 
This includes selecting appropriate daughtercraft for a given objective, assembling spacecraft from modular components, or even manufacturing new assets from onboard or \textit{in situ} resources. 
Each of these choices constitutes a constrained optimization problem under limited resources, particularly when daughtercraft are expendable (e.g., impactors).

Adaptivity is also critical for mission resilience. 
For long-duration missions like ORIGIN, the system must maintain functionality and science return in the presence of anomalies and degradation. 
This includes responding to unexpected failures (e.g., hardware faults or micrometeoroid damage) and accommodating gradual performance decline over time, such as reduced power generation or component aging. 
Robust adaptive strategies are therefore essential to sustain operations in highly uncertain and evolving environments.

\section{Conclusion}
\label{sec:conclusion}

This paper presented the foundations of Planetary Exploration 3.0 (PE~3.0), a new paradigm for robotic space exploration in which radically adaptive, software-defined space systems enable missions to unvisited worlds in the outer Solar System and beyond. As discussed in \Cref{sec:intro}, the transition from the current PE~2.0 paradigm, which relies on the incremental, multi-mission approach successful at Mars, is driven by the realities of exploring distant and poorly characterized environments, where long cruise times and limited prior knowledge render iterative mission campaigns impractical. PE~3.0 replaces inter-mission adaptation with in-mission adaptability, enabled by three key paradigm changes: the shift from hypothesis-driven to capability-driven missions, from hardware-defined to software-defined systems, and from rigid, requirement-based systems engineering to more flexible, capability- and principle-driven approaches. Together, these shifts redefine how missions are conceived, designed, and executed in the presence of deep uncertainty.

\Cref{sec:system} examined the implications of this paradigm for systems engineering. Traditional constructs such as static science traceability matrices and fixed requirements must be augmented or replaced by dynamic, scenario-based formulations that evolve with mission knowledge. Design methods emphasizing flexibility—such as affordance-based design, set-based design, and real options—enable systems to retain latent capabilities and defer commitment under uncertainty. Verification and validation likewise shift from exhaustive requirement compliance toward demonstrating system-level capabilities across representative scenarios. These changes collectively establish a new systems engineering foundation that prioritizes adaptability, resilience, and continuous learning throughout the mission lifecycle.

Sections~\ref{sec:sdss} and \ref{sec:intelligence} addressed the technological and algorithmic enablers of PE~3.0. Software-defined space systems provide the hardware-level flexibility required for adaptation, through reconfigurability, multi-functionality, and modular architectures. However, this flexibility can only be realized with advanced onboard intelligence capable of autonomous decision-making across science operations, navigation, control, and embodied interaction with the environment. The integration of these capabilities enables spacecraft to not only respond to unforeseen conditions but to actively learn, reconfigure, and extend their functionality \textit{in situ}, closing the loop between perception, decision-making, and action.

\Cref{sec:missions} demonstrated how these principles manifest in practice through representative PE~3.0 mission concepts, including a Neptune/Triton smart flyby, an ocean world explorer, and an Oort cloud reconnaissance mission. These concepts illustrate how adaptive systems can unify exploratory and hypothesis-driven science within a single mission, dynamically evolving objectives, architectures, and operations in response to discoveries. 

PE~3.0 challenges long-standing assumptions in planetary exploration, but it also opens the door to fundamentally new capabilities. By embracing adaptability as a core design principle, we can extend exploration to the most distant and unknown regions of our Solar System—and ultimately, beyond.

\section{Acknowledgements}
The findings presented here emerge from the first edition of the Keck Institute for Space Studies (KISS) workshop ``One-shot Outer Solar System Exploration with Adaptive Space Systems'' and were developed further in the second workshop held from April 6 to 10, 2026.
This work is supported in part by the W. M. Keck Institute for Space Studies. Part of this work has been conducted at the Jet Propulsion Laboratory, California Institute of Technology, under contract to NASA. Government sponsorship acknowledged. Copyright 2026. All rights reserved.

\bibliography{references}

\end{document}